\documentclass[%
 reprint,
superscriptaddress,
 amsmath,amssymb,
 aps,
 pra,
floatfix,
]{revtex4-2}
\usepackage[dvisvgm, usenames, dvipsnames]{color}
\usepackage{graphicx}
\usepackage{dcolumn}
\usepackage{bm}
\usepackage{float}
\usepackage{siunitx}
\usepackage{color}
\usepackage{xcolor}
\usepackage{hyperref}

\DeclareMathOperator\erf{erf}

\begin{document}

\preprint{APS/123-QED}

\title{A molecular dynamics framework coupled with smoothed particle hydrodynamics for quantum plasma simulations}

\author{Thomas Campbell}
\affiliation{Department of Physics, Clarendon Laboratory, University of Oxford, Parks Road, Oxford OX1 3PU, UK}

\author{Pontus Svensson}
\affiliation{Department of Physics, Clarendon Laboratory, University of Oxford, Parks Road, Oxford OX1 3PU, UK}

\author{Brett Larder}
\affiliation{Department of Physics, Clarendon Laboratory, University of Oxford, Parks Road, Oxford OX1 3PU, UK}
\affiliation{Machine Discovery Ltd, John Eccles House, Oxford, OX4 4GP, UK}

\author{Daniel Plummer}
\affiliation{Department of Physics, Clarendon Laboratory, University of Oxford, Parks Road, Oxford OX1 3PU, UK}

\author{Sam M. Vinko}
\affiliation{Department of Physics, Clarendon Laboratory, University of Oxford, Parks Road, Oxford OX1 3PU, UK}
\affiliation{Central Laser Facility, STFC Rutherford Appleton Laboratory, Didcot OX11 0QX, UK}

\author{Gianluca Gregori}
\email{gianluca.gregori@physics.ox.ac.uk}
\affiliation{Department of Physics, Clarendon Laboratory, University of Oxford, Parks Road, Oxford OX1 3PU, UK}

\date{\today}

\begin{abstract}
We present a novel scheme for modelling quantum plasmas in the warm dense matter (WDM) regime via a hybrid smoothed particle hydrodynamic - molecular dynamic treatment, here referred to as `Bohm SPH'. This treatment is founded upon Bohm's interpretation of quantum mechanics for partially degenerate fluids, does not apply the Born-Oppenheimer approximation, and is computationally tractable, capable of modelling dynamics over ionic timescales at electronic time resolution. Bohm SPH is also capable of modelling non-Gaussian electron wavefunctions. We present an overview of our methodology, validation tests of the single particle case including the hydrogen 1s wavefunction, and comparisons to simulations of a warm dense hydrogen system  performed with wave packet molecular dynamics.
\end{abstract}

\maketitle

\section{\label{sec:intro}Introduction}

Warm dense matter (WDM) \cite{riley2021warm} is an exotic state of matter transitional between a solid and a plasma, inheriting properties from both. There has been growing interest in the laser-driven production, diagnosis, theoretical treatment, and simulation of WDM in the preceding decades. This has been driven by the advent of high power laser facilities and associated progress in inertial confinement fusion experiments (ICF) \cite{abu2024achievement}, in which the capsule passes through the WDM regime on the route to ignition \cite{hu2018review}, and interest in astrophysical objects in which WDM naturally occurs such as the Jovian (and similar exoplanet) interior \cite{guillot1999interiors,kerley1972equation}, dwarf stars, and neutron star crusts \cite{daligault2009electron}.

WDM is characterised by simultaneously having strongly coupled ions and quantum degenerate electrons. These characteristics make WDM difficult to treat theoretically, with perturbative techniques unreliable. A range of simulation techniques have been developed including effective ion-ion interaction Molecular Dynamics (MD) \cite{mithen2011extent,kahlert2020thermodynamic}, MD with classical electrons interacting via effective pairwise quantum statistical potentials (QSP) \cite{minoo1981temperature,hansen1983thermal,glosli2008molecular,dimonte2008molecular}, Wave Packet Molecular Dynamics (WPMD) \cite{heller1975time,feldmeier1990fermionic,knaup2003wave}, Quantum Hydrodynamics (QHD) \cite{michta2015quantum,moldabekov2018theoretical}, Density Functional Theory coupled to MD (DFT-MD) \cite{white2013orbital,ruter2014ab}, time-dependent Density Functional Theory \cite{ullrich2011time,baczewski2016x}, and Quantum and Path Integral Monte Carlo approaches \cite{hu2011first,militzer2021first,bonitz2020ab}. All with different levels of approximation and computational cost. DFT-MD in particular is applied widely in the WDM regime to compute ion dynamics. However it applies the Born-Oppenheimer approximation, with the electrons treated as an instantaneously adjusting background (adiabatically) and their dynamics not captured.

Dynamic electron behaviour is essential to estimation of system transport properties such as thermal and electrical conductivity, and in the experimental WDM field, essential to interpreting X-ray Thomson scattering which is often used to diagnose plasma conditions \cite{glenzer2009x,fletcher2014observations,poole2022case}. Moreover, explicit electron dynamics may be important to the accuracy of computed ion dynamics in WDM systems, with the first experimental measurements of ion modes in warm dense methane \cite{white2024speed} highlighting the need for accurate ab initio results to corroborate and inform future experiments. Investigation of ion modes in a warm dense Aluminium system in Ref. \cite{mabey2017strong} via a simple Langevin noise model that mimicked the effect of dynamic electrons, suggested that a proper description of dynamic ion - electron and electron - electron interactions is required to predict the ion dynamics accurately. This was supported by further work \cite{yao2021reduced} demonstrating significant difference between DFT-MD results for ion diffusion in warm dense hydrogen with results from the non-adiabatic electron force field (eFF) variant of WPMD \cite{su2009dynamics,angermeier2021investigation}. Latterly this conclusion has been challenged in Ref. \cite{angermeier2023disentangling} performing a like for like comparison of adiabatic and non-adiabatic methodologies via eFF, although uncertainty and limitations remain in the WPMD construction.

WPMD moves beyond the Born-Oppenheimer approximation with equations of motion derived for the electrons via a variational principle \cite{grabowski2014review}. However WPMD's employment of a single Gaussian as each electron's wavefunction can be problematic. At low temperatures in particular, a single Gaussian is too restrictive to produce proper electron screening or resolve the essential atomic physics, or indeed to capture wavefunction break-up \cite{grabowski2013wave}. With a more complete description of the electron state, time-dependent DFT also treats the electron motion explicitly and avoids such restrictive forms for the electron density, but is computationally costly and limited to small particle numbers and short timescales.

Another recent approach to modelling WDM non-adiabatically has been to leverage Bohm's approach to quantum mechanics \cite{bohm1952suggested} (following similar work by de Broglie \cite{de1923quanta} and Madelung \cite{madelung1927quantum}). The reformulation of the single-particle time-dependent Schrödinger equation yields a continuity and momentum evolution equation, with the latter equivalent to that of a classical system but with an additional potential term produced by the kinetic energy operator, the Bohm potential (demonstrated in \ref{sec:Bohm}). The extension of this construction to many-body systems is straightforward (as in section 6 of Ref. \cite{bohm1952suggested}), but calculation of the exact Bohm potential in this case is as complex as solving the exact many-body Schrödinger equation, hence some level of approximation is required. Work by Larder \textit{et al} \cite{larder2019fast} applies a thermally averaged, linearized Bohm potential to capture the quantum kinetic energy of the electrons. This approach applies a two stage methodology where the Bohm potential is first calculated as a function of the equilibrium pair-correlation functions, determined with reference to an ion static structure calculation from an alternative scheme, such as DFT-MD. Once determined, the Bohm potential is then applied in an MD code, equivalent in computational cost to a pairwise classical system.

Here we present a variation of the previous approach for the simulation of WDM: Bohm SPH. In a similar vein to Ref. \cite{larder2019fast}, our platform is non-adiabatic and computationally tractable, able to evolve a warm dense matter system at electronic resolution for ionic timescales. Importantly however, this work moves beyond the two stage methodology and the form of Bohm potential is not restricted to thermal equilibrium. This is accomplished by calculating a many-body Bohm potential on-the-fly with a Smoothed Particle Hydrodynamic (SPH) solver (introduced in the next section) using Gaussian kernels. A further feature of the Bohm SPH construction is access to the continuous spatially resolved electron density. In our methodology we can use multiple SPH particles to model individual electrons. This means that the overall electron shapes are not restricted to the shape of the SPH particles, but can be arbitrarily complex limited only by the number of particles used.

An initial implementation is provided, where further development would entail generalisation of the Coulomb interactions to allow for alternative kernels with more compact support, and more efficient root-finders for determining the optimal SPH kernel scale-lengths. Nonetheless, the method outlined below performs well in tests problems and in particular on a warm dense hydrogen system of the kind that motivates this work.

In section \ref{sec:theory} we outline the theory of the Bohm SPH model. In section \ref{sec:code} we discuss the implementation of Bohm SPH into a molecular dynamics code LAMMPS \cite{thompson2022lammps}, demonstrate its conservation, and highlight its performance in single-particle test problems and scalability in many-body systems. In section \ref{sec:results} we apply the code on a warm dense hydrogen system, and compare the results to those generated via an anisotropic WPMD code, as discussed in Ref. \cite{svensson2023development}.

\section{\label{sec:theory}Theory}

We begin by introducing the SPH methodology, then introduce different force contributions, and finally present the overall Lagrangian solved by Bohm SPH.

Smoothed Particle Hydrodynamics is a meshless scheme for solving fluid equations, applied widely in fields ranging from astrophysics to the computer games industry \cite{monaghan2012smoothed,springel2010smoothed,price2012smoothed}. It obtains approximate numerical solutions of the equations of fluid dynamics by replacing the fluid with a set of particles, whose equations of motion are determined by interpolating from the continuum equations \cite{gingold1977smoothed}. Smoothed Particle Hydrodynamics builds upon the definition of the dirac delta function, defined on a domain \(\Omega\) such that for some continuous function $A(\mathbf{r})$
\begin{eqnarray}
A(\mathbf{r}) =  \int_{\Omega} d\mathbf{r'} A(\mathbf{r'}) \delta(\mathbf{r} - \mathbf{r'}) .
\label{eq:dirac}
\end{eqnarray}
Then by approximating the delta function with a symmetric kernel function \(W\) we can write
\begin{eqnarray}
A(\mathbf{r}) \approx \int_{\Omega} d\mathbf{r'} A(\mathbf{r'}) W(\mathbf{r} - \mathbf{r'},h),
\label{eq:ker}
\end{eqnarray}
where \(h\) is the scale of the kernel function. \(W\) is chosen so that it tends to a delta function in the limit \(h \rightarrow 0\). In the SPH scheme the fluid is divided into small mass particles with mass \(m_{b}\), density \(\rho_{b}\) and position \(\mathbf{r}_{b}\), discretising the integral in equation \eqref{eq:ker} into a summation gives

\begin{eqnarray}
A(\mathbf{r}) = \sum_{b} m_{b} \frac{A_{b}}{\rho_{b}}W(\mathbf{r} - \mathbf{r}_{b},h),
\label{eq:sum}
\end{eqnarray}

\noindent where \(A_{b}\) is the value of the function \(A\) at position \(\mathbf{r}_{b}\). Gradients of the quantity \(A(\mathbf{r})\) can then be calculated similarly,

\begin{eqnarray}
\nabla A(\mathbf{r}) = \sum_{b} m_{b} \frac{A_{b}}{\rho_{b}}\, \nabla W(\mathbf{r} - \mathbf{r}_{b},h).
\label{eq:sum_der}
\end{eqnarray}

In the above \(h\) is a fixed scale length, but can be made into a dynamic per-particle variable. The scale length for particle \(b\), \(h_{b}\), is set according to the local density through the relation

\begin{eqnarray}
h_{b} = \zeta \Big(\frac{m_{b}}{\rho_{b}}\Big)^{\frac{1}{\mathrm{d}}},
\label{eq:denswid}
\end{eqnarray}

\noindent where \(\mathrm{d}\) is the dimension of the system, \(\zeta\) is a constant that must be larger than 1 for stability \cite{ben2000convergence}, and is typically set to approximately 1.3 \cite{monaghan2005smoothed}.  This enforces that the mass in the kernel volume (set by $h_{b}$) is kept constant \cite{springel2010smoothed}, ensuring good neighbour support for each SPH particle.

Clearly, knowledge of the density, \(\rho_{b}\), is needed for any of the previous quantities. The density itself, with dynamic kernel lengths, is computed by

\begin{eqnarray}
\rho_{a} = \sum_{b} m_{b} W(\mathbf{r}_{a} - \mathbf{r}_{b},h_{a}),
\label{eq:dens_base}
\end{eqnarray}

\noindent which is solved alongside equation \eqref{eq:denswid} to determine \(h_{a}\) for each particle, the set \(\{h_{a}\}\). While this is typically done using root-finding procedures such as Newton-Raphson, we have adopted a simple fixed-point iterator which is faster and sufficient to achieve the accuracy we need, as shown later.

As demonstrated in Ref. \cite{price2012smoothed}, the equations of motion for the SPH particles are easily derivable from a discrete version of the continuum Lagrangian of hydrodynamics. We derive those equations here, noting their applicability to a molecular dynamics implementation. Beginning with the continuum Lagrangian,

\begin{eqnarray}
\mathfrak{L} = \int d\mathbf{r}\, \Big[ \frac{\rho \mathbf{v}^{2}}{2} - \rho u(\rho,s)\Big],
\label{eq:contLan}
\end{eqnarray}

\noindent where \(u\) is an internal energy per unit mass and \(\mathbf{v}\) is the velocity. We discretise equation \eqref{eq:contLan} into an SPH form

\begin{eqnarray}
\mathfrak{L} = \sum_{b} \frac{m_{b}}{\rho_{b}} \,\Big[\frac{\rho_{b}\mathbf{v}_{b}^{2}}{2} - \rho_{b} u_{b}(\rho_{b},s_{b})\Big] \\
 = \sum_{b} \, m_{b}\Big[\frac{\mathbf{v}_{b}^{2}}{2} - u_{b}(\rho_{b},s_{b})\Big],
\label{eq:discLan}
\end{eqnarray}

\noindent and assuming this Lagrangian is differentiable, the standard Euler-Lagrange equations follow

\begin{eqnarray}
\frac{d}{dt}\Big(\frac{\partial \mathfrak{L}}{\partial \mathbf{v}_{a}}\Big) - \frac{\partial \mathfrak{L}}{\partial \mathbf{r}_{a}} = 0.
\label{eq:eullang}
\end{eqnarray}

The derivative of the Lagrangian with respect to position \(\partial \mathfrak{L}/ \partial \mathbf{r}_{a}\) is computed by considering the first law of thermodynamics

\begin{eqnarray}
dU = TdS - PdV,
\label{eq:therm1}
\end{eqnarray}

\noindent with \(T\) the temperature, \(S\) the entropy, \(P\) the pressure and \(V\) the volume. Noting that the change in volume can be given by \(dV = -m/\rho^{2} d\rho\), and using per mass quantities, we have

\begin{eqnarray}
du = Tds + \frac{P}{\rho^{2}}d\rho,
\label{eq:therm2}
\end{eqnarray}

\noindent leading to, at constant entropy

\begin{eqnarray}
\frac{\partial \mathfrak{L}}{\partial \mathbf{r}_{a}} =  - \sum_{b} \, m_{b} \left. \frac{\partial u_{b}}{\partial \rho_{b}}\right\vert_{s} \frac{\partial \rho_{b}}{\partial \mathbf{r}_{a}} 
= - \sum_{b} \, m_{b} \frac{P_{b}}{\rho_{b}^{2}} \frac{\partial \rho_{b}}{\partial \mathbf{r}_{a}} .
\label{eq:EulLag}
\end{eqnarray}

The derivative of the density \(\rho_{b}\) with respect to the coordinate \(\mathbf{r}_{a}\) is given by

\begin{eqnarray}
\frac{\partial \rho_{b}}{\partial \mathbf{r}_{a}} = \frac{1}{\Omega_{b}} \sum_{c} m_{c} \left.\frac{\partial W_{bc}(h_{b})}{\partial \mathbf{r}_{a}}\right\vert_{\{h_{b}\}}(\delta_{ba}-\delta_{ca}),
\label{eq:densdynder}
\end{eqnarray}

\noindent where \(W_{bc} (h_{b}) = W(\mathbf{r}_{b} - \mathbf{r}_{c}, h_{b}) \). The derivative term inside the summation keeps the scale lengths \(\{h_{a}\}\) constant, and the term \(\Omega_{b}\) is given by

\begin{equation}
\Omega_{b} = 1 - \frac{\partial h_{b}}{\partial \rho_{b}}\sum_{c}m_{c}\frac{\partial W_{bc}(h_{b})}{\partial h_{b}},
\label{eq:Omega}
\end{equation}

\noindent where from equation \eqref{eq:denswid}

\begin{equation}
\frac{\partial h_{b}}{ \partial \rho_{b}}= - \frac{h_{b}}{\rho_{b}\mathrm{d}}, 
\label{eq:hderrho}
\end{equation}

\noindent with d the number of dimensions.

Combining the above, we eventually arrive at the momentum equation for standard SPH

\begin{equation}
\frac{d \mathbf{v}_{a}}{dt} =  - \sum_{b}m_{b} \Big[ \frac{P_{a}}{\Omega_{a}\rho_{a}^{2}} \left.\frac{\partial W_{ab}(h_{a})}{\partial \mathbf{r}_{a}}\right\vert_{\{h_{b}\}} + \frac{P_{b}}{\Omega_{b}\rho_{b}^{2}} \left.\frac{\partial W_{ab}(h_{b})}{\partial \mathbf{r}_{a}}\right\vert_{\{h_{b}\}}\Big].
\label{eq:sphmom}
\end{equation}

\noindent As we discuss in the next section, by replacing the ordinary pressure with a quantum (Bohm) pressure, we will be able to model the evolution of quantum systems. This has been done previously in Ref. \cite{mocz2015numerical}, applied to a 1d quantum harmonic oscillator, solving the non-linear Schrödinger equation in 2d, and the Gross-Pitaevskii-Poisson equation in 3d. 

The internal energy per particle can also be shown to evolve according to \cite{price2012smoothed}

\begin{eqnarray}
\frac{d u_a}{d t}=\frac{P_a}{\Omega_a \rho_a^2} \sum_b m_b\left(\mathbf{v}_a-\mathbf{v}_b\right) \cdot \left.\frac{\partial W_{ab}(h_{a})}{\partial \mathbf{r}_{a}}\right\vert_{\{h_{b}\}},
\label{eq:internal_energy}
\end{eqnarray}

\noindent where the velocity data must be at the same timestep as the position data (which can be a little cumbersome for typical leapfrog integrators).

The symmetries of the initial discretised Lagrangian, with no dependence on time, as well as translational and rotational invariance, ensure the conservation of energy, linear and angular momentum respectively. This makes the scheme, with equations of motion for each SPH particle based only on the position of neighbours (in their contributions to the estimation of \(\rho_{j}\) and hence the Bohm pressure \(P_{j}\)), ideal for solving within a molecular dynamic framework. The overall Lagrangian solved, including additional forces, is introduced at the end of the section.

\subsection{\label{sec:Bohm}Bohm Potential}

The Bohm potential \cite{bohm1952suggested} can be derived by using a polar (Madelung \cite{madelung1927quantum}) form of the wavefunction, here demonstrated for a single particle

\begin{eqnarray}
\psi(\mathbf{r},t) = R(\mathbf{r},t)\,\exp\Big[\frac{iS(\mathbf{r},t)}{\hbar}\Big],
\label{eq:polar}
\end{eqnarray}
where \(R\) and \(S\) are real, and \(\mathbf{r}\) is the position vector. The time dependent Schr\"{o}dinger equation, for a particle of mass $m$ under an external potential $V_{ext}$ and with $\hbar$ the reduced Planck's constant,
\begin{eqnarray}
i\hbar \frac{\partial\psi}{\partial t} = - \frac{\hbar^{2}}{2m}\nabla^{2}\psi + V_{ext}\psi,
\label{eq:tdse}
\end{eqnarray}
yields with this polar form of \(\psi\), by equating the imaginary and real components, equations for \(R\) and \(S\) respectively
\begin{eqnarray}\label{R}
\frac{\partial R}{\partial t} = - \frac{1}{2m}[R\nabla^{2}S + 2\nabla R\cdot \nabla S]
\label{eq:Rev}
\end{eqnarray}
\begin{eqnarray}\label{HJ}
\frac{\partial S}{\partial t} = - \Big[\frac{(\nabla S)^{2}}{2m} + V_{ext} - \frac{\hbar^{2}}{2m}\frac{\nabla^{2}R}{R}\Big].
\label{eq:Sev}
\end{eqnarray}

We can write \(R = n^{\frac{1}{2}}\) where here \(n(\mathbf{r},t)\) is the probability density of the particle in phase space, so we can express equation \eqref{eq:Rev} as, 
\begin{eqnarray}
\frac{\partial n}{\partial t} + \nabla \cdot\Big(n \frac{\nabla S}{m}\Big) = 0,
\label{eq:prob}
\end{eqnarray}

\noindent which is a probability conservation equation where \( \frac{\nabla S}{m} \) gives the velocity. In an SPH scheme, the discretisation of the continuity equation is simply given by the density summation of equation \eqref{eq:dens_base}, where normalised kernel functions \(\int_{V} W dV = 1\) ensure a conservation of total mass.

Importantly we recognise that equation \eqref{eq:Sev} is the classical Hamilton Jacobi equation with an additional quantum potential, the Bohm potential
\begin{eqnarray}\label{eq:Bohm}
V_{B}(\mathbf{r},t) = - \frac{\hbar^{2}}{2m}\frac{\nabla^2{R(\mathbf{r},t)}}{R(\mathbf{r},t)}.
\end{eqnarray}

We require the many-body form of the Bohm potential for treating quantum plasmas. Following Ref. \cite{gregori2019modified}, the \(N\)-body Bohm potential can be written as 

\begin{equation}
    V_{B}^{(N)} = -\frac{\hbar^{2}}{2m}\sum_{i}^{N} \frac{\nabla_{i}^{2}|\psi|}{|\psi|},
\end{equation}
where \(\psi = \psi(\mathbf{r}_{1},\mathbf{r}_{2},...,\mathbf{r}_{N})\) is the \(N\)-body wavefunction, and \(\nabla_{i}\) is the gradient with respect to the \(i\)th particle coordinates. For computational feasibility we now derive the Quantum Hydrodynamic (QHD) form of the Bohm potential \cite{moldabekov2018theoretical,bonitz2019quantum,wyatt2005quantum}, which is a function only of the total density of the electron fluid.
Taking a Hartree product for the many-body wavefunction \(\psi(\mathbf{r}_{1},\mathbf{r}_{2},...,\mathbf{r}_{N}) = \phi_{1}(\mathbf{r}_{1})\phi_{2}(\mathbf{r}_{2})...\phi_{N}(\mathbf{r}_{N})\), where \(\phi_{i}(\mathbf{r}_{i})\) is the \(i\)th particle wavefunction, the expectation value of the Bohm potential is 

\begin{equation}\label{many1}
\begin{split}
    \langle V_{B} \rangle & = \int d\mathbf{r}_{1}\int d\mathbf{r}_{2}...\int d\mathbf{r}_{N} V_{B}^{(N)}|\psi|^{2} \\
    & = \int d\mathbf{r} \sum_{i}^{N} V_{i}^{(N)}(\mathbf{r})|\phi_{i}(\mathbf{r})|^{2},
\end{split}
\end{equation}
where
\begin{equation}
    V_{i}^{(N)}(\mathbf{r}) = - \frac{\hbar^{2}}{2m}\frac{\nabla^{2} |\phi_{i}(\mathbf{r})|}{|\phi_{i}(\mathbf{r})|},
\end{equation}
is the single particle Bohm potential. We note that the Hartree product form of the many-body wavefunction is not antisymmetric, but address this shortcoming with an additional potential to capture Pauli exclusion, as shown in section \ref{sec:symmetry}. The total particle number density is \(n(\mathbf{r}) = \sum_{i} |\phi_{i}|^2 = \sum_{i} n_{i}\), where \(n_{i}\) is the probability distribution for the \(i\)th particle, thus
\begin{equation}\label{many2}
\begin{split}
    \langle V_{B} \rangle  & = \int d\mathbf{r} \sum_{i}^{N} n_{i}\Big(-\frac{\hbar^2}{2m}\frac{\nabla^2 \sqrt{n_{i}}}{\sqrt{n_{i}}}\Big) \\
    & \approx \int d\mathbf{r} \, n(\mathbf{r})\Big(-\gamma \frac{\hbar^2}{2m}\frac{\nabla^2 \sqrt{n}}{\sqrt{n}}\Big),
\end{split}
\end{equation}

\noindent where in the last step we have applied the linearization approximation of QHD, which is exact when all the single particle wavefunction amplitudes are identical \cite{manfredi2001self, manfredi2005model, manfredi2021fluid}, and introduces a linearization constant for fermions, \(\gamma\). Thus we are left with the QHD form for the Bohm potential, as a function of the total number density, a single spatial coordinate

\begin{equation}
    V_{\si{QHD}}(\mathbf{r}) = -\gamma \frac{ \hbar^2}{2m}\frac{\nabla^2 \sqrt{n(\mathbf{r})}}{\sqrt{n(\mathbf{r})}}.
\end{equation}

The linearization constant is equal to \(1\) for bosons, and for fermions in the low temperature limit generally equal to \(1/9\) \cite{moldabekov2015statically,michta2015quantum}, but, by comparison with the limits of the Random Phase Approximation polarization function \cite{moldabekov2018theoretical}, can differ according to wavenumber and frequency. The low frequency and long wavelength limit in particular has additional temperature and density dependencies, with \(\gamma\) ranging from 1/9 at zero temperature increasing up to 1/3 at \(\theta > 1\). However at high frequencies \(> \hbar k^{2}/2m_{e}\), setting \(\gamma = 1\) yields the expected plasmon dispersion relation. In this work, where we are resolving the electron dynamics at sub-attosecond resolution, we apply the high frequency limit of \(\gamma = 1\).

We apply the quantum pressure tensor form, as in Ref. \cite{mocz2015numerical}, used in the equation of motion for SPH particles (equation \eqref{eq:EulLag})

\begin{equation}
    P_{B}(\mathbf{r}) = -\frac{\hbar^2}{4m} n \, \nabla \otimes \nabla \ln n,
    \label{eq:presssimple}
\end{equation}

\noindent where \(\otimes\) is the outer product, and which is related to the Bohm potential via \cite{manfredi2021fluid}

\begin{equation}
    \nabla \cdot P_{B} = n \nabla V_{B}.
\end{equation}

The pressure tensor is symmetric. We can expand  equation \eqref{eq:presssimple} for the \(xy\) value as an example

\begin{equation}
\begin{split}
P_{B_{xy}} & = -\frac{\hbar^2}{4m} n\, \partial_{x} \Big[\partial_{y} \ln (n) \Big] \\
& = -\frac{\hbar^2}{4m} n\, \partial_{x} \Big[\frac{\partial_{y}n}{n}  \Big] \\
& = \frac{\hbar^2}{4m}\, \Big[ \frac{\partial_{x}n \, \partial_{y}n}{n} - \partial_{xy}n\Big].
\end{split}
\end{equation}

\noindent The Bohm pressure expression is calculated in an SPH discretisation. We use the same as in Ref. \cite{mocz2015numerical}, but with difference terms in both the first and second derivatives of the density, selected following the conservation analysis discussed in section \ref{sec:conserve}. The Bohm pressure for the \(xy\) component of the \(i\)th SPH particle is

\begin{equation}
    P_{B_{i,xy}} = \frac{\hbar^2}{4m} \sum_{j} \, \frac{m_{j}}{\rho_{j}} \Big[\frac{\partial_{x}n_{j} \, \partial_{y}n_{j}}{n_{j}} - \partial_{xy}n_{j} \Big] \, W_{ij}(h_{i}),
    \label{eq:presstens}
\end{equation}

\noindent where \(W_{ij}(h_{i}) = W(\mathbf{r}_{i} - \mathbf{r}_{j},h_{i})\). The equations of motion are also discretised the same way as in Ref. \cite{mocz2015numerical}, namely, for the \(x\) component

\begin{equation}
\begin{aligned}
\frac{d v_i^x}{d t}=-\sum_j m_j\Big\{& \frac{\left[P_{B_{i, x x}}, P_{B_{i, x y}}, P_{B_{i, x z}}\right]}{\rho_i^2 \Omega_i} \cdot \left.\frac{\partial W_{ij}(h_{i})}{\partial \mathbf{r}_{i}}\right\vert_{\{h_{b}\}}\\
+ & \frac{\left[P_{B_{j, x x}}, P_{B_{j, x y}}, P_{B_{j, x z}}\right]}{\rho_j^2 \Omega_j} \cdot \left.\frac{\partial W_{ij}(h_{j})}{\partial \mathbf{r}_{i}}\right\vert_{\{h_{b}\}}\Big\}.
\end{aligned}
\end{equation}

Finally, given that the Bohm pressure is not a scalar, the equation for the evolution of the per-particle internal energy \eqref{eq:internal_energy} has to be modified. We use the following expression

\begin{equation}
\frac{d u_{B_a}}{d t}=\frac{1}{\Omega_a \rho_a^2} \sum_b m_b [P_{B_a} \left(\mathbf{v}_a-\mathbf{v}_b\right)] \cdot \left.\frac{\partial W_{ab}(h_{a})}{\partial \mathbf{r}_{a}}\right\vert_{\{h_{b}\}},
\label{eq:internal_energy_bohm}
\end{equation}

\noindent where \(u_{B_a}\) is the internal energy due to the Bohm pressure for the \(a\)th particle.

We note that an improvement to implementing the QHD-level Bohm pressure tensor would be to compute the Bohm pressure forces on density distributions belonging to each individual electron in the system. This `Many-Fermion' Bohm potential, as discussed in \cite{moldabekov2022towards}, was investigated but initial tests indicated that its computational cost was prohibitive, hence the QHD Bohm term is the focus of this work.

Having introduced SPH and the Bohm potential, we can discuss the general construction of the model. Bohm SPH uses the Smoothed Particle Hydrodynamic solver to calculate the Bohm force, where the electron density is modelled by Gaussian SPH particles. The density distribution of the SPH particles is taken to be the charge distribution and used to directly calculate the Coulomb potential which couples the electronic component with point ions. This smearing of the electrons prevents asymptotic ion-electron Coulomb attraction, similar to the wave packets in WPMD being the electron charge density, and somewhat similar to the diffractive form of Quantum Statistical Potentials (QSP), such as the Kelbg Potential \cite{kelbg1963theorie,filinov2003improved}. Although the resolution of the SPH distribution is controlled numerically by the kernel sizes and not a de Broglie type scale length as in QSP. In order to resolve better the electron density we run simulations with more SPH particles than electrons \(N_{S} > N_{e}\). When doing so, the overall mass and charge density of the system is kept consistent, as well as the charge to mass ratio of SPH particles. We apply confining potentials in this case to localise individual electrons and put the velocities of their centres of mass into a target distribution. This avoids unphysical thermal effects caused by the additional degrees of freedom, discussed at greater length in section \ref{sec:thermalisation}.

\subsection{\label{sec:coul}Coulomb Forces}

A central step in our hybrid SPH-MD modelling of the electrons comes in the treatment of the Coulomb interaction. We take the kernel used to interpolate the density and Bohm pressure as the real charge density distribution of each particle. We have adopted a Gaussian kernel function for \(W\) because of its readily integrable form, and derived the exact SPH Coulomb equations of motion. A more general treatment of the Coulomb interaction would follow the SPH treatment of softened gravity of Ref. \cite{price2007energy}, without recourse to a specific kernel function. We leave this to future work.

The Gaussian charge density profile is

\begin{eqnarray}
\rho_{e_{j}}(\mathbf{r}) = n_{j}(\mathbf{r})q_{j} = \frac{q_{j}}{(\pi h_{j}^2)^{3/2}} \exp\Big(-\frac{|\mathbf{r}-\mathbf{r}_{j}|^{2}}{h_{j}^{2}}\Big),
\label{eq:gauss}
\end{eqnarray}

\noindent with \(q_{j}\), \(\mathbf{r}_{j}\), and \(h_{j}\) its fractional charge, centre of mass, and scale length (width) respectively. The Coulomb potential between an SPH particle and an ion can then be calculated by the analytic integral

\begin{eqnarray}
V_{ij} = \int d\mathbf{r} \frac{Ze}{4\pi \epsilon_{0}|\mathbf{r} - \mathbf{r}_{i}|} \rho_{e_{j}}(\mathbf{r}) ,
\label{eq:coul_SPH_ie_int}
\end{eqnarray}

\noindent where \(\mathbf{r}_{i}\) is the position of the ion, and \(Z\) its charge, yielding with \(r_{ij} = |\mathbf{r}_{i} - \mathbf{r}_{j}|\),

\begin{eqnarray}
V_{ij} = \frac{Z e q_{j}}{4\pi \epsilon_{0}r_{ij}} \erf \Big(\frac{r_{ij}}{h_{j}}\Big).
\label{eq:coul_SPH_ie}
\end{eqnarray}

\noindent The integral in \eqref{eq:coul_SPH_ie_int} assumes a simple point ion with \(1/r\) Coulomb interaction. This could also be changed to a pseudopotential to include the effect of core electrons, the form of which can be Gaussian-decomposed to enable analytical solutions.

The procedure for the pairwise SPH particle Coulomb potential is similar, integrating across both Gaussian charge clouds

\begin{eqnarray}
V_{jk} = \int \int d\mathbf{r} d\mathbf{r'} \frac{\rho_{e_{j}}(\mathbf{r})\rho_{e_{k}}(\mathbf{r'})}{4\pi \epsilon_{0}|\mathbf{r} - \mathbf{r'}|} ,
\label{eq:coul_SPH_ee_int}
\end{eqnarray}

\noindent yielding for particles \(j\) and \(k\) 

\begin{eqnarray}
V_{jk} = \frac{q_{j}q_{k}}{4\pi \epsilon_{0}r_{jk}} \erf \Big(\frac{r_{jk}}{\sqrt{h_{j}^{2} + h_{k}^{2}}}\Big).
\label{eq:coul_SPH_ee}
\end{eqnarray}

\noindent When using dynamic kernel widths, these pairwise potentials actually become many-body, via the particle width \(h_{j}\) dependence on the local density in equation \eqref{eq:denswid}

\begin{eqnarray}
\frac{\partial V_{jk}}{\partial \mathbf{r}_{l}} = \left.\frac{\partial V_{jk}}{\partial \mathbf{r}_{l}}\right\vert_{\{h\}} + \sum_{m} \left.\frac{\partial V_{jk}}{\partial h_{m}} \right\vert_{\mathbf{r}_{l}} \frac{\partial h_{m}}{\partial \mathbf{r}_{l}}.
\label{eq:manybodyder}
\end{eqnarray}

It is instructive to expand this expression to the exact form implemented within Bohm SPH. Starting with the first term on the right hand side of \eqref{eq:manybodyder}, which is non-zero only for \(l = j\) or \(l=k\), and defining \(h_{j}^{2} + h_{k}^{2} = M_{jk}^{2}\)

\begin{eqnarray}
\begin{split}
\left.\frac{\partial V_{jk}}{\partial \mathbf{r}_{j}}\right\vert_{\{h\}} = \frac{\xi_{jk}}{ r_{jk}}\Big[\left(\frac{2}{\sqrt{\pi} M_{jk}}\right) & \exp\left(-\frac{r_{jk}^{2}}{M_{jk}^{2}}\right) \\
- \frac{1}{r_{jk}} & \erf \left(\frac{r_{jk}}{ M_{jk}} \right) \Big] \hat{\mathbf{r}}_{jk},
\label{eq:dr_exp}
\end{split}
\end{eqnarray}

\noindent where \(\xi_{jk} = q_{j}q_{k}/4 \pi \epsilon_0\) and \(\mathbf{r}_{jk} = \mathbf{r}_{j} - \mathbf{r}_{k}\). The second term on the right of equation \eqref{eq:manybodyder} can be expanded via the chain rule as

\begin{eqnarray}
 \sum_{m} \left.\frac{\partial V_{jk}}{\partial h_{m}} \right\vert_{\mathbf{r}_{l}} \frac{\partial h_{m}}{\partial \mathbf{r}_{l}} = \sum_{m} \left.\frac{\partial V_{jk}}{\partial h_{m}} \right\vert_{\mathbf{r}_{l}} \frac{\partial h_{m}}{\partial \rho_{m}}\frac{\partial \rho_{m}}{\partial \mathbf{r}_{l}}.
\label{eq:chaindynker}
\end{eqnarray}

\noindent The derivative of the pairwise Coulomb potential with respect to kernel scale length \(h_m\) is

\begin{eqnarray}
\left.\frac{\partial V_{jk}}{\partial h_m}\right\vert_{\mathbf{r}_{l}} = - \frac{2\xi_{jm}h_{m}}{\sqrt{\pi}M_{mj}^{3}}\exp\left(-\frac{r_{mj}^{2}}{M_{mj}^{2}}\right).
\label{eq:coulderwid}
\end{eqnarray}

\noindent Then, the remaining terms follow from the SPH treatment of dynamic kernel lengths and the definition of density, as in equations \eqref{eq:densdynder} and \eqref{eq:hderrho}. We can insert these expressions into an equation for the total electronic Coulomb force acting on an SPH particle

\begin{equation}
\mathbf{F}_{a}^{ee} = -\frac{\partial \si{V}_{C}^{ee}}{\partial \mathbf{r}_{a}} = - \frac{\partial}{\partial \mathbf{r}_{a}} \sum_{j\neq k}  \frac{\xi_{jk}}{r_{jk}} \erf \left(\frac{r_{jk}}{M_{jk}} \right),
\label{eq:totalCoulombder}
\end{equation}

\noindent where \(\si{V}_{C}^{ee}\) indicates the total Coulomb potential between SPH particles in the system. After some algebra this can be expressed in compressed form as 

\begin{eqnarray}
\mathbf{F}_{a}^{ee} = \Gamma_a - \theta_{a} \left.\frac{\partial \rho_{a}}{\partial \mathbf{r}_{a}}\right\vert_{\{h\}} - \sum_{k \neq a} \theta_{k} m_{a}\left.\frac{\partial W_{ak}(h_{k})}{\partial \mathbf{r}_{a}}\right\vert_{\{h\}}
\label{eq:elecoulforce}
\end{eqnarray}

\noindent where the respective terms are

\begin{equation}
\begin{split}
\Gamma_b = \sum_{k \neq b } \frac{\xi_{bk}}{r_{bk}} \Big[\frac{1}{r_{bk}} & \erf \left(\frac{r_{bk}}{M_{bk}} \right)  \\
- \left(\frac{2}{\sqrt{\pi} M_{bk}}\right) & \exp\left(-\frac{r_{bk}^{2}}{M_{bk}}\right)\Big] \hat{\mathbf{r}}_{bk},
\end{split}
\end{equation}

\begin{equation}
\theta_{b} = \sum_{k \neq b} \frac{2h_{b}^{2}\xi_{bk}}{3 \rho_{b} \Omega_{b}\sqrt{\pi}M_{bk}^{3}} \exp\left(-\frac{r_{bk}^{2}}{M_{bk}^{2}}\right).
\end{equation}

\noindent Note that the pairwise cutoff for computing \(\Gamma\) and \(\theta\) are different. \(\Gamma\) has the same cutoff as the Coulomb potential, whereas \(\theta\) must have the same cutoff as that used for the computation of SPH quantities. This is expected given that the \(\theta \) forces originate from neighbouring SPH particle coordinates determining kernel scale lengths through equations \eqref{eq:dens_base} and \eqref{eq:denswid}.

The full Coulomb force term for an electronic SPH particle interacting with ions, starting from the potential equation \eqref{eq:coul_SPH_ie}, is almost exactly equivalent, but with \(M_{bk}\) being replaced by \(h_b\) and one of the SPH charges replaced by the ion charge \(Z_{i}e\). Notably however, the Coulomb force acting on the ion by the electronic component is only given by the \(\Gamma\) term above, since the ion coordinates do not determine the dynamic SPH scale lengths. The conservative nature of these expressions are demonstrated in section \ref{sec:conserve}.

\subsection{\label{sec:symmetry}Symmetry Effects}

When dealing with a many-fermion system indistinguishable particles cannot exist in the same state. Construction of the QHD Bohm potential is ignorant of this requirement so we must include symmetry effects via an additional potential. Having focused on implementation of the Bohm term in the first iteration of this model rather than highly accurate exchange effects, we include exchange effects in a simple manner by borrowing a spin-averaged symmetry potential from QSP, which we denote \(V_{P}\) for Pauli exclusion. Precisely we employ the temperature dependent equation derived in \cite{minoo1981temperature} and subsequently applied in MD simulations of thermal relaxation such as \cite{hansen1983thermal,glosli2008molecular}

\begin{eqnarray}
V_{P} = k_{B} T \ln(2) \, \exp\Big[-\frac{1}{\ln(2)}\Big(\frac{r}{\lambda_{ee}}\Big)^{2}\Big],
\label{eq:minoo}
\end{eqnarray}

\noindent where \(\lambda_{ee} = \hbar \Big(k_{B}Tm_{e}\Big)^{-1/2}\). The target temperature is used in equation \eqref{eq:minoo} rather than the instantaneous temperature.  Taking the derivative of the Pauli potential between the \(i\)th and \(j\)th particle, where \(\mathbf{r}_{ij} = \mathbf{r}_{i} - \mathbf{r}_{j}\) yields the force

\begin{equation}
\mathbf{F}_{P_{i}} = -\frac{\partial V_{P_{ij}}}{\partial \mathbf{r}_{i}} = 2 k_{B}T \left(\frac{\mathbf{r}_{ij}}{\lambda_{ee}^{2}}\right) \, \exp\left[-\frac{1}{\ln(2)}\left(\frac{r}{\lambda_{ee}}\right)^{2}\right].
\label{eq:minoo_force}
\end{equation}

When using sub-electron resolution in the model, with \(N_{ppe}\) SPH particles per electron, the interaction is scaled by \(1/N_{ppe}^{2}\), and interactions between `same-electron' particles are removed. This conserves the total Pauli potential in the system and, if same-electron particles are on top of one another, replicates the pairwise electron interaction (\(N_{ppe} = 1\)). This factor naturally appears in the SPH discretisation of the Pauli potential, as shown later.

Additionally, the inclusion of the repulsive Minoo potential in the model helps protect against the tensile instability of SPH \cite{morris1996analysis,monaghan2000sph,price2012smoothed}, which can otherwise result in an unphysical clustering of SPH particles. Indeed, we have not observed the tensile instability in applications of Bohm SPH thus far.

\subsection{\label{sec:resolution}SPH Resolution}

A feature of Bohm SPH is the ability to resolve the electronic component with arbitrary resolution, dependent only on the number of SPH particles used. A useful metric for determining whether the charge density is well resolved is comparison of the average kernel width \(\overline{h}\) to the expected screening length of the plasma \(\lambda_{S}\), we desire \(\overline{h} < \lambda_{S}\). In the classical and quantum limits the relevant screening lengths will be the Debye \(\lambda_{D}\) and Thomas Fermi \(\lambda_{TF}\) lengths respectively. We use equation 6 of Ref. \cite{glenzer2009x} to define the screening length \(\lambda_{S}\), which returns \(\lambda_{D}\) and \(\lambda_{TF}\) in the appropriate limits

\begin{eqnarray}
\lambda_{S}^{-2} = \kappa_{e}^{2} = \frac{n_{e}e^{2}}{\epsilon_{0}k_{B}T_{e}} \frac{F_{-1/2}(\eta_{e})}{F_{1/2}(\eta_{e})},
\label{eq:screening}
\end{eqnarray}

\noindent where \(\eta_{e}\) is the dimensionless chemical potential \(\mu_{e}/k_{B}T_{e}\), and \(F_{\nu}\) denotes a Fermi integral of order \(\nu\).

The requirement of good neighbour support for SPH schemes \cite{ben2000convergence} means that we cannot arbitrarily reduce the kernel widths of the particles. Instead, we increase the number of particles. For the remainder of this manuscript, when discussing systems with \(N_{ppe}\) particles per electron, we have scaled all SPH particle masses and charges by \(1/N_{ppe}\) to ensure the correct mass and charge density. Via equation \eqref{eq:denswid}, we can define the average kernel width \(\overline{h}\) for a system with electron density \(n_{e}\) 

\begin{eqnarray}
\overline{h} = \zeta(N_{ppe}n_{e})^{-1/3}.
\label{eq:dens_wid_ave}
\end{eqnarray}

\begin{figure}[h]
\includegraphics[width=0.48\textwidth]{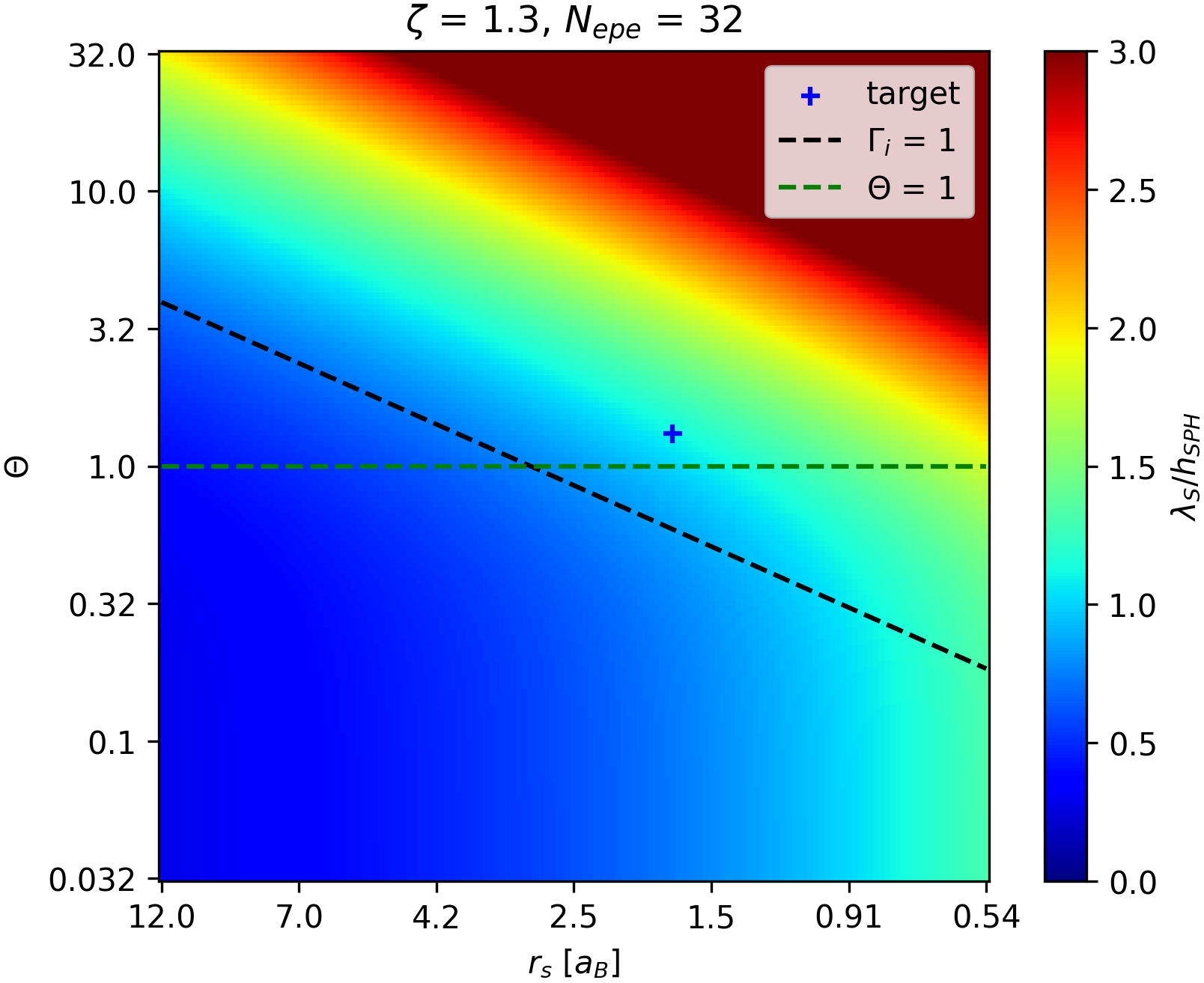}
\caption{\label{fig:kernel_map} \small Ratio of the screening length \(\lambda_{S}\) to the average SPH kernel width  \(\overline{h}\) for ionised hydrogen with \(N_{ppe} = 32\). The `target' system is investigated in section \ref{sec:results}.}
\end{figure}

Figure \ref{fig:kernel_map} demonstrates that we require \(N_{ppe} = 32\) when setting \(\zeta = 1.3\) to resolve the warm dense hydrogen system investigated in \ref{sec:results} with Wigner Seitz radius \(r_{s} = (3/4 \pi n_{e})^{1/3}  = 1.75 \, \si{a_{B}}\) and degeneracy parameter \(\theta = k_{B}T/E_{F} = 1.32\), with \(\si{a_{B}}\) the Bohr radius and \(E_{F}\) the Fermi energy. 

\subsection{\label{sec:thermalisation}Confinement Potential}

We must consider the implication of the ion and electron systems having the same temperature since we are not employing the Born-Oppenheimer approximation. All SPH particles are degrees of freedom and hence contribute to the thermal energy. We generally require \(N_{S} > N_{e}\) to sufficiently resolve the electron density according to  \(\overline{h} < \lambda_{S}\), so the Bohm SPH system will have additional thermal energy compared to the physical one, as demonstrated by equipartition

\begin{eqnarray}
\sum_{i}^{N} \frac{1}{2} m_{i} \langle \mathbf{v}_{i}^{2}\rangle = \frac{3}{2} N k_{B} T,
\label{eq:equipart}
\end{eqnarray}

\noindent where \(T\) is the target temperature of the system.  We assume all \(N_{S} = N_{e}N_{ppe}\) SPH particles (with \(N_{e}\) the number of electrons) have identical masses \(m_{s}\) and move at an average speed \(\overline{v}\) given by

\begin{eqnarray}
\frac{N_{S} m_{s} \overline{v}^{2}}{2} = \frac{3}{2} N_{S} k_{B} T.
\label{eq:equipart2}
\end{eqnarray}

\noindent Since we have \(N_{ppe}\) SPH particles per electron, the SPH particle mass scales as \(m_{s} = m_{e}/N_{ppe}\) to ensure the correct mass density, so we rewrite \eqref{eq:equipart2} as

\begin{eqnarray}
\frac{m_{e} \overline{v}^{2}}{2N_{ppe}} = \frac{3}{2} k_{B} T.
\label{eq:equipart3}
\end{eqnarray}

\noindent Rearrangement of \eqref{eq:equipart3} yields

\begin{eqnarray}
\overline{v} = \sqrt{\frac{3 N_{ppe} k_{B} T}{m_{e}}},
\label{eq:equipart4}
\end{eqnarray}

\noindent demonstrating how the average, and indeed the thermal, speed of the SPH particles scales proportionally to \(\sqrt{N_{ppe}}\), causing an unphysical Bohm-Gross dispersion and a spurious ion screening.

This problem is not unique to Bohm SPH. In fact it also appears in Particle-in-Cell (PIC) simulations. There, the temperatures of charge macroparticles are typically scaled by the macroparticle weight to address unphysical velocities \cite{dawson1983particle,acciarri2024should}. In our case we cannot apply a general scaling as the ions in our model are not treated identically to the electrons, but as point-particles whose temperature must be fixed at \(T\).

One approach for addressing this problem would be to model the ions and electrons under separate thermostats, with ions at \(T\) and electrons at \(T/N_{ppe}\). This can be problematic for collecting reliable ion trajectories as large values of \(N_{ppe}\) demand a strong thermostat to prevent the ions equilibriating with the SPH bulk.

An alternative approach, used in this work, is to introduce a quadratic confining potential to localise individual electrons and to apply a thermostat to their centres of mass (CoM) which are subsequently released into an NVE (microcanonical) ensemble. After equilibriating these centres of mass at the target temperature, plasmon data computed from their trajectories then avoids the numerical Bohm-Gross dispersion mentioned above. Furthermore, the trajectory data is collected while the whole system is in NVE rather than the ionic and electronic components being maintained at separate temperatures.

SPH particles are allocated a parent electron and forced toward their centre of mass via the potential

\begin{eqnarray}
V_{C}(\mathbf{r}_{i}) = g|\mathbf{r}_{i} - \mathbf{R}|^{2},
\label{eq:confine}
\end{eqnarray}

\noindent where \(g\) is varied to adjust the size of the parent electron, \(\mathbf{r}_{i}\) is the position of a target particle, and the centre of mass \(\mathbf{R} = \sum_{j} \mathbf{r}_{j} / N_{ppe}\) for equal mass particles. In a system with periodic boundary conditions, the centre of mass is calculated according to the formulation of Ref. \cite{bai2008calculating}. The confinement force on a particle is given by

\begin{equation}
\mathbf{F_{C_{i}}} = - \frac{\partial \sum_{j}V_{C_{j}}(\mathbf{r_{j}})}{\partial \mathbf{r}_{i}},
\label{eq:confine_force1}
\end{equation}

\noindent where the sum is over the \(N_{ppe}\) SPH particles which belong to the same centre of mass. This can be expanded as

\begin{equation}
\mathbf{F_{C_{i}}} = -\frac{2g}{N_{ppe}} \left[(\mathbf{r}_{i} - \mathbf{R})\left(N_{ppe}-1\right) - \sum_{j \neq i} (\mathbf{r}_{j} - \mathbf{R})\right].
\label{eq:confine_force2}
\end{equation}

\noindent In a box without periodic boundary conditions, this simply reduces to

\begin{equation}
\mathbf{F_{C_{i}}} = 2g(\mathbf{R} - \mathbf{r}_{i}),
\label{eq:confine_force3}
\end{equation}

\noindent but with periodic boundary conditions, the vectors \(\mathbf{r}_{a} - \mathbf{R}\) need to be computed using the correct (closest) projection of \(\mathbf{R}\), which can be the overall centre of mass plus or minus a box length in all principal directions \(\mathbf{R} \pm L \hat{\mathbf{i}}\). This does not necessarily reduce to equation \eqref{eq:confine_force3}.

As stated earlier, we remove the repulsive Coulomb and Pauli potentials between particles belonging to the same electron, while retaining the Bohm interaction. We perform a scan of \(g\) values when comparing outputs from Bohm SPH to anisotropic WPMD in section \ref{sec:results}.

\subsection{\label{sec:Lagrangian}Full Lagrangian}

It is instructive to consider the full Lagrangian of the Bohm SPH model. Using the interactions listed above, we can define a Lagrangian for a quantum plasma system with electron density \(n(\mathbf{r})\) and \(N_{I}\) point ions. To start, we include self interactions and omit the confining potential

\begin{equation}
\begin{split}
\mathfrak{L} = \sum_{i=1}^{N_{I}} \Big[\frac{1}{2}M_{i}\mathbf{v}_{i}^{2} - \sum_{j>i}^{N_{I}} \frac{(Ze)^{2}}{4\pi \epsilon_{0}|\mathbf{r}_{i} - \mathbf{r}_{j} |}\Big] \\ +  \int d\mathbf{r} \, n(\mathbf{r})\Big\{\frac{1}{2}m_{e}\mathbf{v}(\mathbf{r})^{2}
- m_{e}u_{B}(\mathbf{r}) 
- \sum_{i}^{N_{i}} \frac{Ze^{2}}{4\pi \epsilon_{0}|\mathbf{r}_{i} - \mathbf{r} |} \\
- \int d\mathbf{r'} \, \frac{n(\mathbf{r'})}{2} \Big[\frac{e^{2}}{4\pi \epsilon_{0}|\mathbf{r'} - \mathbf{r} |} + V_{P}(|\mathbf{r'} - \mathbf{r}|)\Big] \Big\}
\label{eq:Lag1}
\end{split}
\end{equation}

\noindent where \(M_{I}\) is the ion mass, \(Z\) its ionisation, \(u_{B}\) the internal Bohm energy per unit mass (whose derivative with respect to density is related to the Bohm pressure tensor via equation \eqref{eq:therm2}), and \(n\) the number density of electrons, with a factor of \(1/2\) included in the second integral to prevent double counting. Now, for the electron kinetic, Bohm, and Pauli terms, we apply the SPH discretisation, while for the Coulomb interactions we integrate exactly using the charge density distribution given by the SPH Gaussian kernels. This procedure eliminates all the integral terms, replacing them with summations that can be implemented into a molecular dynamics structure. Furthermore, we remove the Coulomb and Pauli interactions between SPH particles belonging to the same parent electron and, if enabled, introduce confining potentials for each electron. With \(N_{S}\) SPH particles we have

\begin{equation}
\begin{split}
\mathfrak{L} = \sum_{i=1}^{N_{I}} \Big(\frac{1}{2}M_{i}\mathbf{v}_{i}^{2} -\sum_{j>i}^{N_{I}}\frac{(Ze)^{2}}{4\pi \epsilon_{0}r_{ij}} \Big) \\ + \sum_{a=1}^{N_{S}} \Big\{\frac{1}{2}m_{a}\mathbf{v}_{a}^{2} - m_{a}u_{B_{a}} 
- \sum_{i=1}^{N_{I}} \Big[\frac{Zeq_{a}}{4\pi \epsilon_{0}r_{ia}} \erf \Big(\frac{r_{ia}}{\sqrt{2h_{a}^{2}}}\Big)\Big]  \\
- \sum_{b=1}^{N_{S}'}  \frac{1}{2} \Big[\frac{q_{a}q_{b}}{4\pi \epsilon_{0}r_{ab}} \erf \Big(\frac{r_{ab}}{\sqrt{2(h_{a}^{2} + h_{b}^{2}))}}\Big) + \frac{1}{N_{ppe}^{2}}V_{P}(r_{ab})\Big] \Big\} \\ 
  - \sum_{c=1}^{N_{e}} \sum_{d=1}^{N_{ppe}} \Big[g |\mathbf{r}_{d} - \mathbf{R}_{c}|^{2}\Big].
\label{eq:Lag3}
\end{split}
\end{equation}

\noindent Here the SPH variables have subscript \(a\) and \(b\), with \(m_{a}\) the SPH particle mass, \(q_{a}\) its fractional charge, \(h_{a} = h_{a}(\rho_{a})\) its dynamic kernel width, and \(r_{ab} = |\mathbf{r}_{a} - \mathbf{r}_{b}|\). \(N_{S}'\) indicates that particles \(b\) belonging to the same electron as particle \(a\) are excluded, and where the index \(c\) runs over \(N_{e}\) whole electrons and \(d\) over \(N_{ppe}\) members of each electron.

\section{\label{sec:code}Numerics}

Bohm SPH has been implemented via modification of LAMMPS, an open source classical molecular dynamics code with a focus on materials modeling \cite{thompson2022lammps}. This includes routines for the Bohm, Pauli, real-space Coulomb interactions (compatible with the Ewald decomposition \cite{deserno1998mesh}), a fixed point iterator for computing kernel widths from local densities, confining potentials compatible with Periodic Boundary Conditions \cite{bai2008calculating}, as well as a Nosé-Hoover thermostat  \cite{evans1985nose} that operates on the electron centres of mass rather than the SPH particles. Simulations are performed using a velocity-Verlet integrator.

\subsection{\label{sec:conserve}Conservation}

Following the equations of motion for SPH particles of the previous section, we demonstrate their conservative properties. First, we begin by comparing various forms for the density derivatives used in the Bohm pressure tensor on a simple periodic box interacting only through the Bohm pressure force. We list the mass densities here, but note that number densities are required in the Bohm expressions, simply related by \(n = \rho/m_{e}\) with \(m_{e}\) the electron mass in the appropriate unit. It is well established in the SPH method that naïve derivatives of equation \eqref{eq:sum}, as in equation \eqref{eq:sum_der}, are not the most accurate \cite{fatehi2011error,korzilius2016improved,basa2009robustness}, in fact various alternative expressions exist for SPH derivatives. The forms and associated names investigated here are, for the first derivative

\begin{align}
\textrm{basic:}\,\,\, & \partial_{x} \rho_{i} = \sum_{j} m_{j} \, \partial_{x}W_{ij}(h_{i}) \label{eq:basic1} \\
\textrm{F2:}\,\,\, & \partial_{x} \rho_{i} = \sum_{j} m_{j} \, \left(1 - \frac{\rho_{i}}{\rho_{j}}\right) \,\partial_{x}W_{ij}(h_{i}), \label{eq:F2}
\end{align}

\noindent where `F2' is borrowed as a label from Ref. \cite{fatehi2011error}, and for the second derivative

\begin{align}
\textrm{basic:}\,\,\, & \partial_{xy} \rho_{i} = \sum_{j} m_{j} \, \partial_{xy}W_{ij}(h_{i}) \label{eq:basic2} \\
\textrm{Mocz:}\,\,\, & \partial_{xy} \rho_{i} = \sum_{j} m_{j} \, \left(1 - \frac{\rho_{i}}{\rho_{j}}\right) \,\partial_{xy}W_{ij}(h_{i}) \label{eq:mocz2}\\
\textrm{D1:}\,\,\, & \partial_{xy} \rho_{i} = \sum_{j} \frac{m_{j}}{\rho_{j}} \, \partial_{y} \rho_{j} \,\partial_{x}W_{ij}(h_{i}) \label{eq:doub}\\
\textrm{D2:}\,\,\, & \partial_{xy} \rho_{i} = \sum_{j} \frac{m_{j}}{\rho_{j}} \, \left(\partial_{y} \rho_{j} -\partial_{y} \rho_{i}\right)\,\partial_{x}W_{ij}(h_{i}), \label{eq:doubdiff}
\end{align}
 
\noindent where `Mocz' corresponds to the form used in Ref. \cite{mocz2015numerical}. It is worth noting that the second derivatives D1 and D2 are not symmetric \(\partial_{xy} \neq \partial_{yx}\), meaning the Bohm pressure tensor is no longer symmetric. They also require an additional loop over neighbours to compute the first derivative terms.

The eight combinations of derivatives are used on an 1024 SPH particle system, with total mass equivalent to 16 electrons in a cubic box of length \(7.11 \, \mathrm{a_{B}}\), the same density as the electrons in the warm dense hydrogen system in the following section. All simulations are initialised identically from rest with timestep \(0.1 \, \mathrm{as}\), \(\zeta = 1.3\) and cutoff \(3\,\overline{h}\). The energy outputs are plotted in figure \ref{fig:bohm_energy}. All combinations of derivatives conserve momentum to machine precision, as shown in appendix figure \ref{fig:bohm_mom}. The energy conservation, using equation \eqref{eq:internal_energy_bohm} to compute the rate of change of the internal energy and accumulate it for each particle over time, is less uniform. A summary of the energy conservation is given in table \ref{tab:bohm_compare}, where the drift of the total energy is squared and summed over the 5 fs duration of the simulation as an indicator of energy conservation.

\begin{figure}[h]
\includegraphics[width=0.48\textwidth]{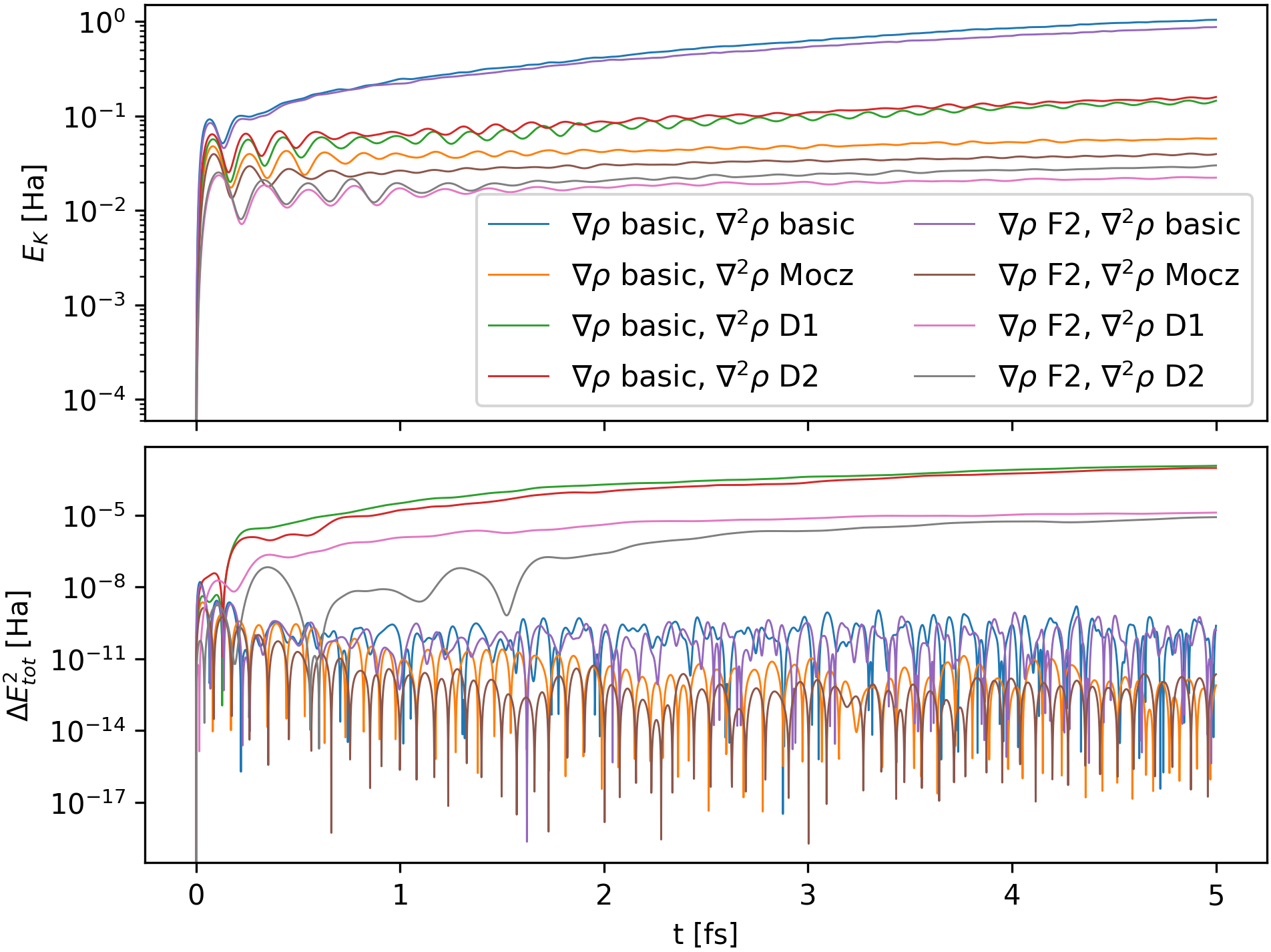}
\caption{\label{fig:bohm_energy}\small Kinetic energy and the square of the total energy drift of a Bohm-only system with different derivative combinations.}
\end{figure}

\begin{table}
\caption{\label{tab:bohm_compare}} \small Energy conservation of derivative combinations for a test Bohm-only simulation with periodic boundary conditions, sorted from best to worst, all to 3sf.
\begin{tabular}[t]{|c|c|c|c|}
\hline
$\nabla \rho$ & $\nabla^{2} \rho$ & $\int \Delta E_{tot}^{2} dt \,[\mathrm{Ha}^{2}\,\mathrm{fs} ] $  \\[1.5ex]
\hline 
\textbf{F2} & \textbf{Mocz} & $\mathbf{1.01\cdot10^{-11}} $\\
basic & Mocz &  $2.05\cdot10^{-11} $\\
F2 & basic &  $9.62\cdot10^{-11} $\\
basic & basic &  $1.29\cdot10^{-10} $\\
F2 & D2 & $1.13\cdot10^{-6} $ \\
F2 & D1 & $2.89\cdot10^{-6} $ \\
basic & D2 & 0.000136\\
basic & D1 & 0.000191\\
\hline

\end{tabular}
\end{table}

The combination of F2 for the first and Mocz for the second derivative seems to be optimal for computing the Bohm pressure tensor in a system with periodic boundary conditions, having the best energy conservation. Derivatives D1 and D2 have surprisingly poor conservation, which we speculate may be related to the breaking of symmetry in the second derivatives.

Next we investigate the conservation of the Coulomb SPH force expressions on a one-component-plasma (OCP) system. The periodic box has 5248 particles, corresponding to 82 electrons, with length \(12.261 \, \mathrm{a_{B}}\), once again at the same density of electrons in the warm dense hydrogen system of the following section. All particles interact with one another via long-range Coulomb potentials. For comparison, both the SPH Coulomb interaction and the standard point Coulomb interaction are examined, with identical initialisations of a random particle arrangement. Specifically, the standard LAMMPS interaction coul/long is used for the points. The simulations are run with a timestep of \(0.5 \, \mathrm{as}\), \(\zeta = 1.3\), a real-space cutoff of \(r_{c} = 5.5 \,\mathrm{a_{B}}\), and a ewald parameter equal to \(3/r_{c}\). The energy outputs are shown in figure \ref{fig:coul_energy}. SPH Coulomb conserves momentum to machine precision as shown in appendix figure \ref{fig:coul_mom}. After an initial jump, the total energy of the SPH Coulomb system oscillates with similar amplitude to that of the point Coulomb, and the kinetic energy of the SPH Coulomb system equilibrates at a lower value than the point Coulomb as expected. This validates the conservation of the novel SPH Coulomb interaction.

\begin{figure}[h]
\includegraphics[width=0.48\textwidth]{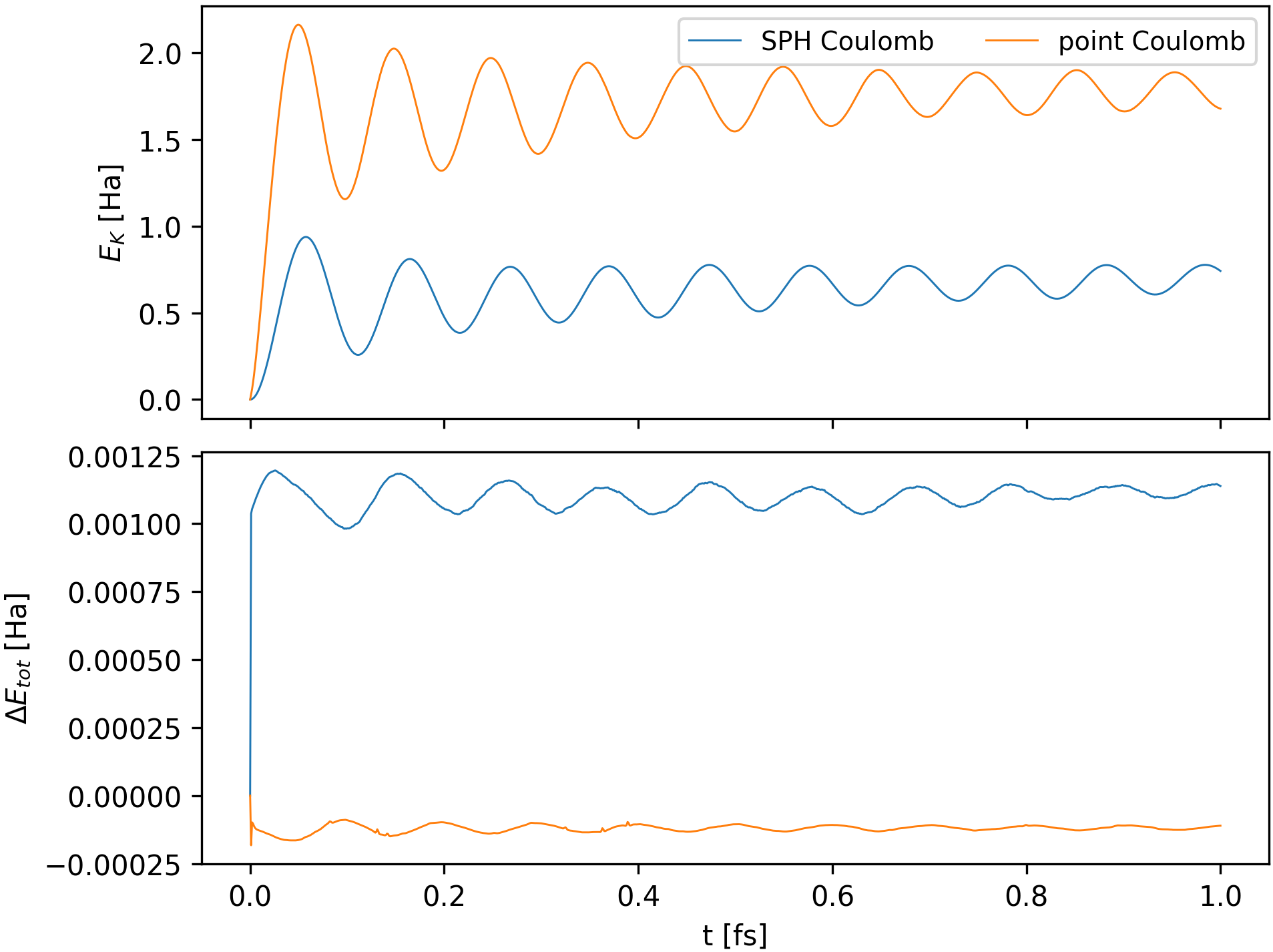}
\caption{\label{fig:coul_energy}\small Kinetic energy and drift of the total energy drift of a test OCP Coulomb system comparing standard point and SPH Coulomb interactions.}
\end{figure}

\subsection{\label{sec:scaling}Scaling}

\begin{figure*}[!ht]
\includegraphics[width=0.98\textwidth]{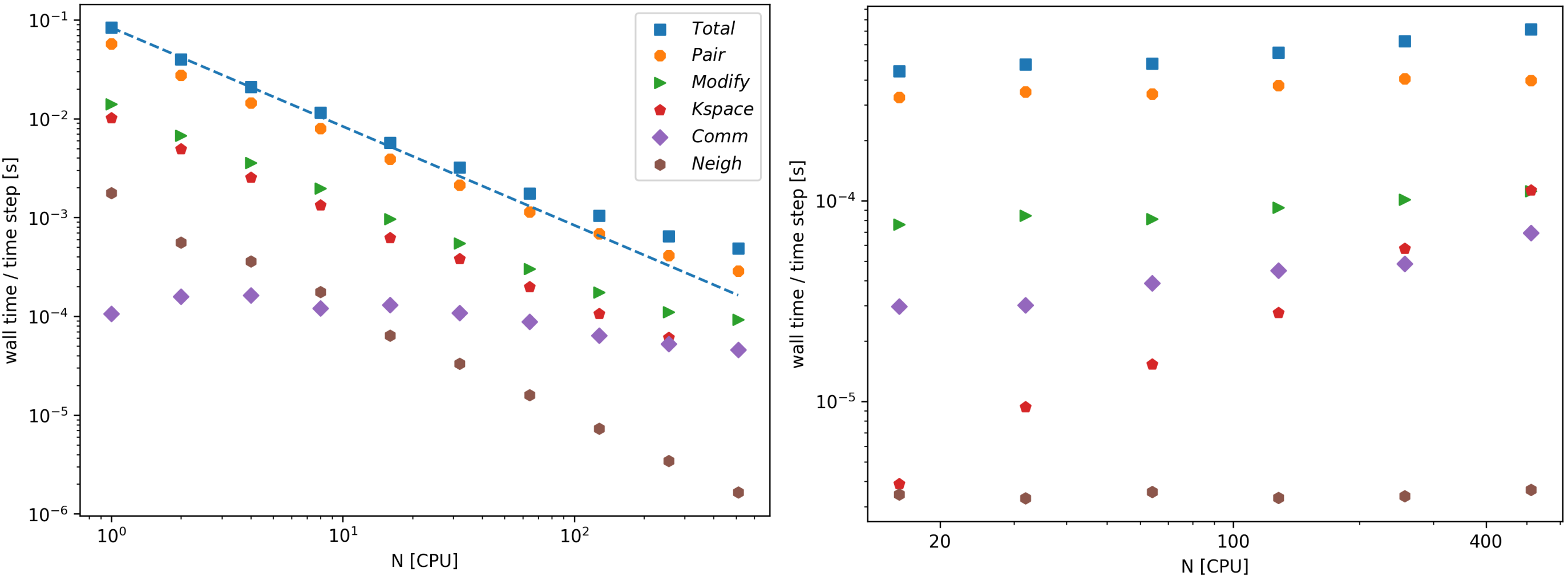}
\caption{\label{fig:StrongWeakScale}\small Strong (left) scaling of a warm dense hydrogen system with 512 protons and 16384 SPH particles, and weak (right) scaling of the same warm dense hydrogen system with a resolution of \(N_{ppe} = 32\) SPH particles per electron. Includes runtime statistics for individual LAMMPS modules. Perfect scaling is indicated by the dashed blue line. The individual module contributions are: the real space force computation in `Pair', the dynamic electron width and centre of mass computation within `Modify', Ewald Coulomb calculation in `Kspace', neighbour list construction in `Neigh', and communication times between MPI processors in `Comm'.} 
\end{figure*}

The bespoke SPH module, utilising the LAMMPS framework, has excellent parallel scaling. Our module is separate to one previously implemented in LAMMPS (see Ref. \cite{ganzenmuller2011implementation}). The strong scaling of the warm dense hydrogen system investigated in section \ref{sec:results}, at a density of \(n_{e} = 3.006 \, \si{g}/\si{cm^{3}}\) and temperature \(T = 21.54 \, \si{eV}\), with 512 protons and 16384 SPH electron particles (\(N_{ppe} = 32\)), with all interactions computed (Coulomb, Bohm, Pauli and Confinement) is demonstrated in the left inset of figure \ref{fig:StrongWeakScale}. The scaling contributions from modules within LAMMPS are also plotted alongside the total time. Perfect scaling is given by the relation 

\begin{equation}
t_{N} = t_{1}/N,
\label{eq:scaling}
\end{equation}

\noindent where \(t_{N}\) is the wall time per timestep for a simulation running on \(N\) processors. We see in figure \ref{fig:StrongWeakScale} that in the example warm dense hydrogen system, the compute time only begins to notably diverge from perfect scaling at around 100 CPU. This divergence is also dependent on the system size and cutoff radii values for the various force interactions, and hence can be tuned with variation of these parameters.

The weak scaling of Bohm SPH is presented in the right inset of figure \ref{fig:StrongWeakScale}, with consistent compute times observed across the number of processors. The weak scaling is computed with the resolution kept constant and the box size increased. As shown in figure \ref{fig:StrongWeakScale}, the most computationally intensive parts of Bohm SPH are `Pair' and `Modify'. The SPH Bohm pressure force and Coulomb forces comprise the majority of the `Pair' compute time, while the calculation of the densities, dynamic kernel scale lengths and centres of mass comprise almost all the `Modify' compute time. Using an alternative kernel with better compact support and therefore smaller neighbour cutoffs, such as the cubic spline \cite{schoenberg1946contributions}, would speed up future implementations.

\subsection {\label{sec:BohmConfine}Oscillator Ground State}

In the ground state tests of the oscillator and hydrogen, we did not use F2 \eqref{eq:F2} for the first order density derivatives as we found it caused greater instability than a basic derivative \eqref{eq:basic1} in these particular cases that have a free boundary. SPH schemes generally require special care to handle free boundaries \cite{monaghan2005smoothed}. We have not taken such care due to our systems of interest being continuous plasmas treated with periodic boundary conditions. Despite this, Bohm SPH demonstrates good agreement on two single particle problems which have analytical solutions: the ground states of the 3d quantum harmonic oscillator and the hydrogen atom.

To validate the Bohm expressions used, we first investigate a reduced system interacting only via the Bohm pressure force and a quadratic confining potential. Unlike in many-electron simulations the confining potential here is centred on a fixed coordinate rather than the centre of mass of the SPH distribution. Running simulations with \(N_{S} = 256\) SPH particles and dynamic kernel widths we damp the system to zero temperature to achieve the ground state of a quantum harmonic oscillator. In this single wavefunction example, the Bohm equations are exact. Taking a Gaussian probability density profile as shown below, equating the expectation energies of the confining potential and the Bohm potential gives a simple relation between the confining potential strength \(g\) and the wavefunction width \(H\). The Gaussian ground state density distribution is

\begin{eqnarray}
n(\mathbf{r}) = |\psi(\mathbf{r})|^{2} = \frac{1}{(\pi H^{2})^{3/2}}\exp \Big(-\frac{|\mathbf{r}|^{2}}{H^{2}}\Big),
\label{eq:oscground}
\end{eqnarray}

\noindent where \(H\) is the overall width of the wavefunction. Here the confining potential is centred on the origin, and has expectation energy

\begin{eqnarray}
\langle V_{c} \rangle = \int d\mathbf{r} \, g r^{2} |\psi(\mathbf{r})|^{2} = \frac{3H^{2}g}{2},
\label{eq:confexp}
\end{eqnarray}

\noindent and the expectation of the Bohm potential

\begin{equation}
\begin{split}
\langle V_{B} \rangle =\int d\mathbf{r} \, \Big(- \frac{\hbar^{2}}{2m} \frac{ \nabla^{2} \sqrt{n(\mathbf{r})}}{\sqrt{n(\mathbf{r})}} \,\Big) |\psi(\mathbf{r})|^{2} \\ 
= \int d\mathbf{r} \, \frac{\hbar^{2}}{2mH^{2}}\Big(3- \frac{r^{2}}{H^{2}}\Big) |\psi(\mathbf{r})|^{2} = \frac{3\hbar^{2}}{4mH^{2}},
\label{eq:Bohmexp}
\end{split}
\end{equation}

\noindent then equating \eqref{eq:confexp} and \eqref{eq:Bohmexp} yields,

\begin{eqnarray}
H = \left(\frac{\hbar^{2}}{2mg}\right)^{1/4}.
\label{eq:bohmconf}
\end{eqnarray}

\begin{figure}[h]
\includegraphics[width=0.48\textwidth]{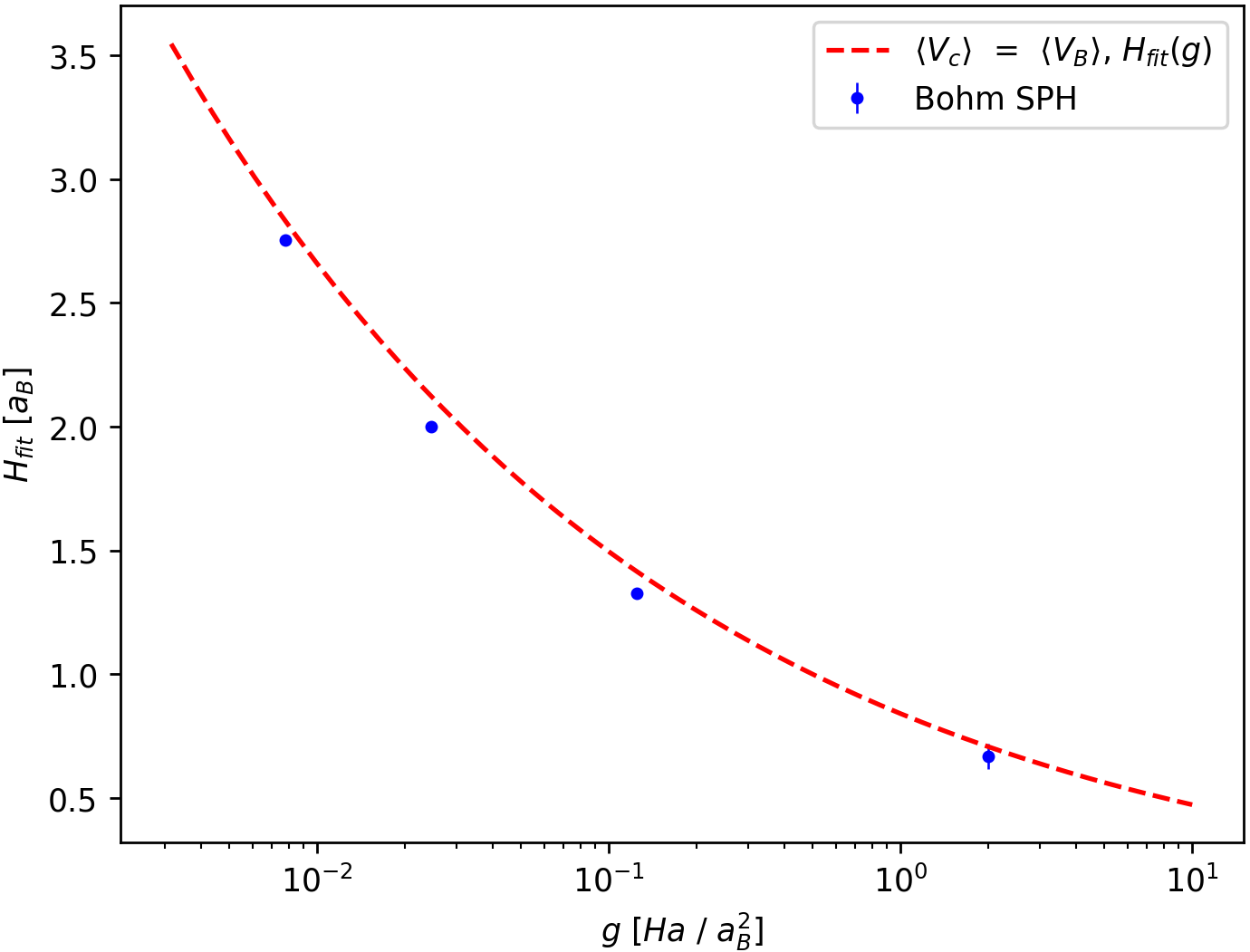}
\caption{\label{fig:BohmConfine}\small Fitted Gaussian width outputs from reduced Bohm SPH simulations of a damped quantum harmonic oscillator compared to expected relation (equation \eqref{eq:bohmconf}). Plotted error is the standard deviation of the width calculations of the final 200 time steps (\(50 \, \si{as}\)) of each run, only visible in the strongest confinement point.}
\end{figure}

After damping, the particles are released into an NVE ensemble to check the stability of the solution and the density distributions are fitted to a Gaussian. The fitted Gaussian widths from four simulations sampling different confining strengths \(g\) are summarised in figure \ref{fig:BohmConfine}, and show excellent agreement with the expected relation \eqref{eq:bohmconf}, validating the implementation of the Bohm pressure tensor.

\subsection {\label{sec:HydGround}Hydrogen Ground State }

Now we further test our implementation of the Bohm and the Coulomb forces by attempting to solve for the ground state of hydrogen. For this single electron system, we do not include the Pauli interaction, Coulomb potentials between SPH particles, or the confining potential. While the ground state of the harmonic oscillator is straightforward to solve in Bohm SPH and relatively insensitive to initial distribution and damping strength, the ground state of hydrogen is more challenging. It is difficult to fully suppress the kinetic energy of the SPH particles. We attribute this to the strength of attraction between electron SPH particles and the central ion (equation \eqref{eq:coul_SPH_ie}) being not only a function of radial separation, but also of the dynamic kernel widths which are dependent on the many-body distribution. 

\begin{figure*}[t]
\includegraphics[width=1.0\textwidth]{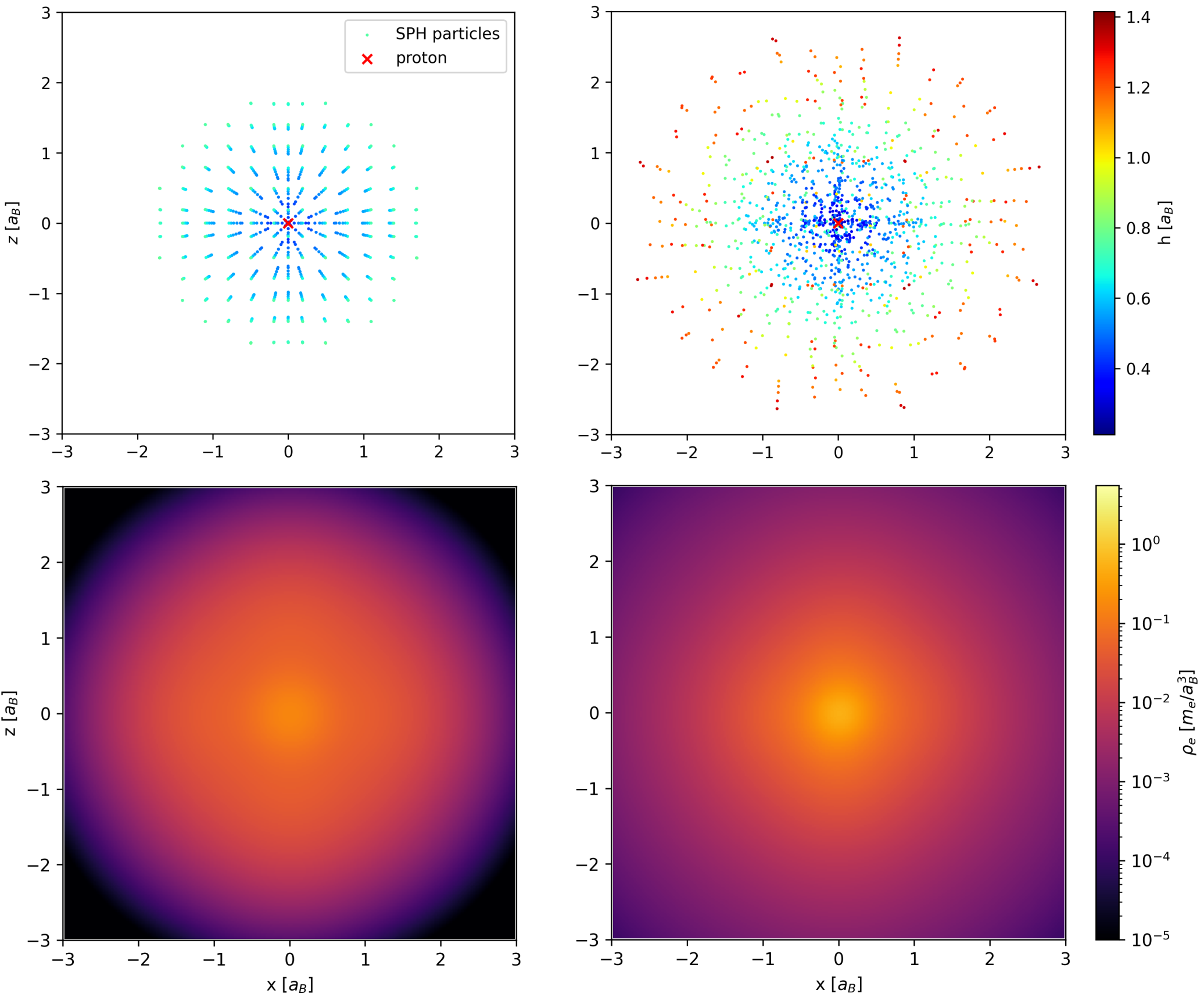}
\caption{\small Initial (left) and final (\(t = 1.44 \, \si{fs}\), right) SPH particle distributions for damped Bohm SPH simulation of the hydrogen ground state with particles initialised within a spherical cutoff \(r_{0} = 2.0 \, \si{a_{B}}\) of the proton. SPH particle position and width (\(h\)) information (top) and continuous density profile of cross section at \(y = 0\) (bottom).}
\label{fig:Hground_array}
\end{figure*}

SPH particles are first initialised on a simple cubic grid around the proton. Comparisons of different starting grids used \(N_{S} = 1237\) SPH particles. The cubic grid terminates within spherical limits to give the system rough initial spherical symmetry. Three initial cutoffs were investigated, \(r_{0} = 2.0 \, \si{a_{B}}\), \(2.5 \, \si{a_{B}}\), and \(3.0 \, \si{a_{B}}\), with lattice parameters of \(0.3 \, \si{a_{B}}\), \(0.375 \, \si{a_{B}}\), and \(0.45 \, \si{a_{B}}\) respectively. The particles are randomly displaced off the grid points prior to running by \(0.005 \, \si{a_{B}}\) to break the exact symmetry. The simulations are all then run with a time step of \(5\times 10^{-4} \, \si{as}\) with a frictional damping term applied, of strength \(1\times 10^{-4} \,\si{Ha} \cdot \si{fs} / \si{a_{B}^{2}}\). A value of \(\zeta = 1.3\cdot\sqrt{2}\) is used to produce larger kernels and promote stability. The initial and final distribution of SPH particles (projected in two dimensions) is shown in figure \ref{fig:Hground_array} for initial cutoff radius \(r_{0} = 2.0 \, \si{a_{B}}\). An additional simulation with \(N_{S} = 2469\) on a grid with lattice parameter of \(0.3 \, \si{a_{B}}\) which terminated within a radius of \(r = 2.5 \, \si{a_{B}}\) was also run to demonstrate convergence of the density and overall energy toward the exact wavefunction solution.

\begin{figure}[h]
\includegraphics[width=0.48\textwidth]{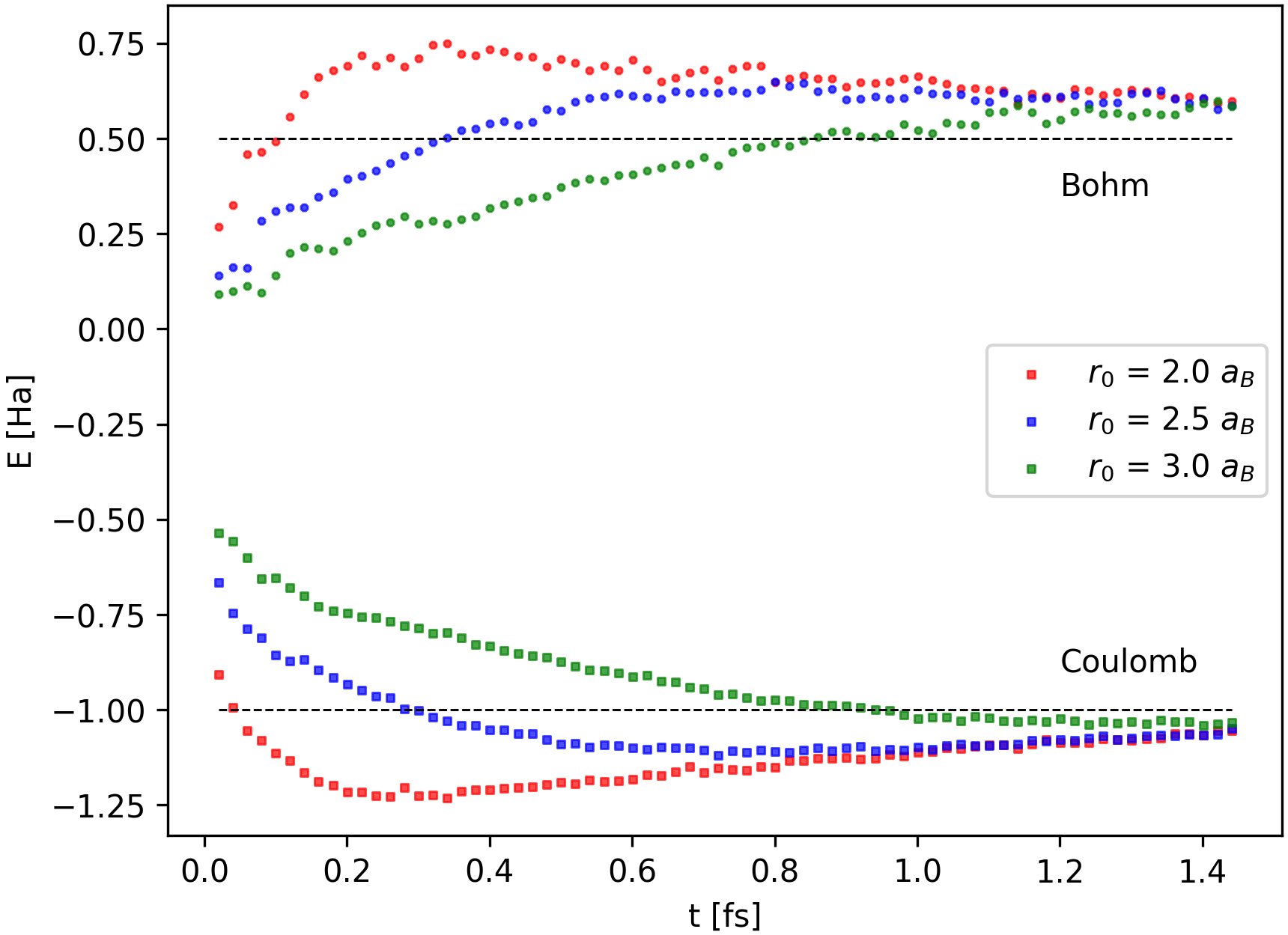}
\caption{\label{fig:HydEnergy}\small Energy evolution of damped \(N_{S} = 1237\) Bohm SPH simulations of hydrogen ground state with initial radii \(r_{0}\) = 2.0, 2.5, and 3.0 \(\si{a_{B}}\), computed from single snapshots of SPH distribution at 0.02 \(\si{fs}\) intervals. Squares indicate Coulomb energy, and circles the Bohm energy. Horizontal lines are the exact 1s wavefunction energies.}
\end{figure}

\begin{figure}[h]
\includegraphics[width=0.48\textwidth]{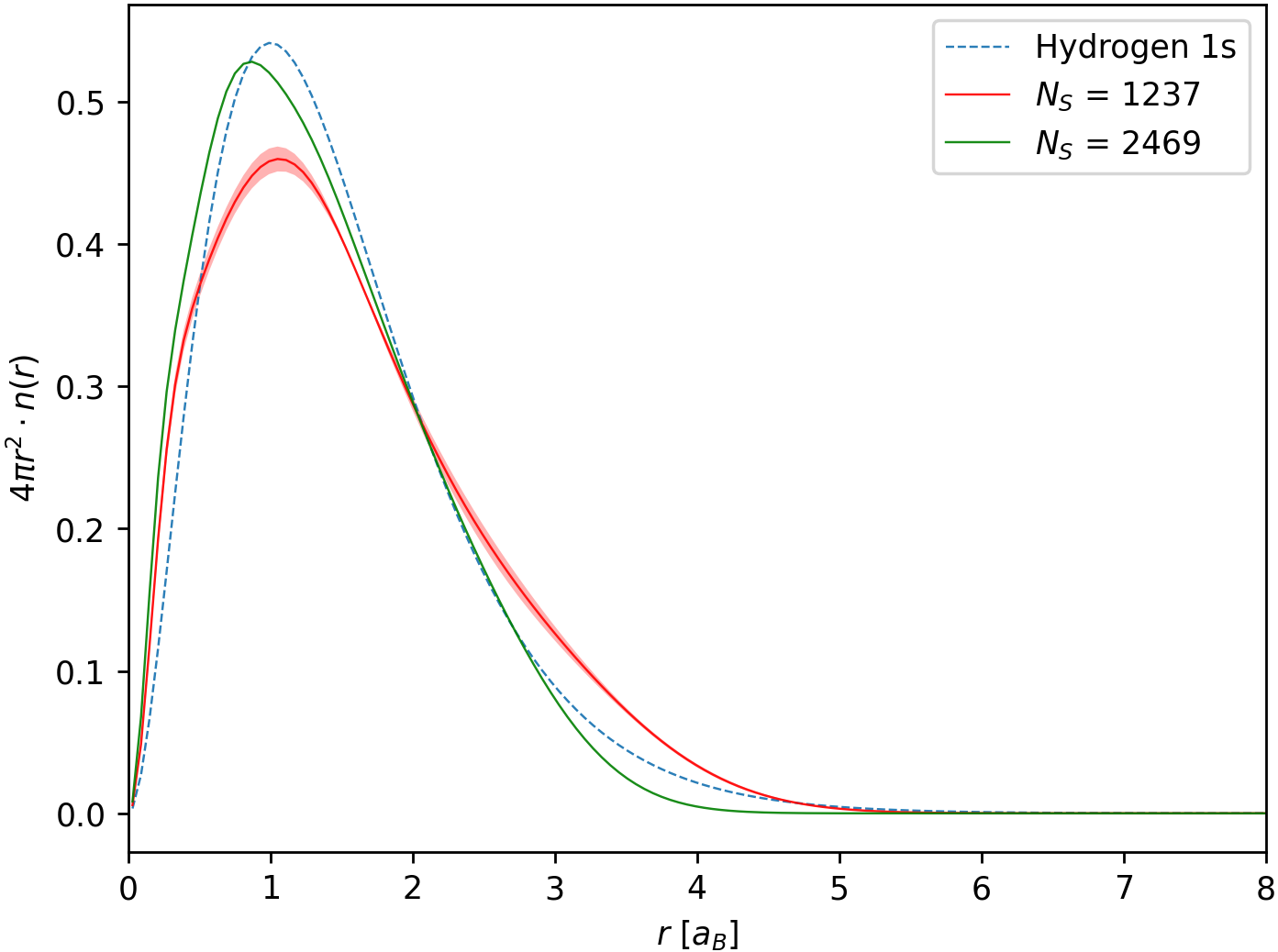}%
\caption{\label{fig:HydGroundAvePlot} \small Density distribution from average of final five snapshots of damped \(N_{S} = 1237\) and \(N_{S} = 2469\) particle Bohm SPH simulations of the hydrogen ground state, compared to exact hydrogen 1s distribution. \(N_{S} = 1237\) average includes values from all three different initial radii, whereas \(N_{S} = 2469\) from a single run with initial radius \(2.5 \, \si{a_B}\). The average total energy of the \(N_{S} = 1237\) results is \(\langle V_{\si{Total}} \rangle = -0.46 \pm 0.01 \, \si{Ha} \), and for \(N_{S} = 2469\) \(\langle V_{\si{Total}} \rangle = -0.513 \pm 0.003 \, \si{Ha}\). Central solid line is mean, with error bar \(\pm\) the standard deviation.}
\end{figure}

The evolution of the separate \(N_{S} = 1237\) Bohm SPH runs is shown in figure \ref{fig:HydEnergy}, which demonstrates each run converging on similar Bohm and Coulomb energies. The average distribution of SPH particles across all three runs in the final 5 snapshots, from \(t = 1.36\) to \(1.44 \, \si{fs}\) at \(0.02 \, \si{fs}\) intervals, is plotted in \ref{fig:HydGroundAvePlot}. The energy averages and errors are given in table \ref{tab:HydGround}. For reference, the best fit (energy) of a single Gaussian to the hydrogen 1s density distribution, of width \(H=1.33 \, a_{B}\), is also included in the table. The Bohm potential is calculated for each SPH particle via the equation 

\begin{eqnarray}
V_{B_{a}} = -\frac{\hbar^{2}}{8m_{e}} \Big[\frac{2 \nabla^{2} n_{a}}{n_{a}} - \frac{(\nabla n_{a})^{2}}{n_{a}^{2}}\Big]
\label{eq:bohmpot}
\end{eqnarray}

\noindent where as discussed the first derivatives of the density are computed without the difference term as in equation \eqref{eq:basic1}, and the second as in equation \eqref{eq:mocz2}. The total Bohm energy of the system is then \(\langle V_{\si{Bohm}} \rangle = \sum_{a} V_{B_{a}} / N_{S}\).

Bohm SPH returns return a total energy value closer to the exact 1s expectation of -0.5 Hartree than the best fit single Gaussian. The convergence of the separate Bohm SPH runs toward a shared ground state, with a more accurate overall energy than the best fit single Gaussian case, validates our treatment of the Coulomb interaction which applies the SPH kernels as real charge distributions.

\renewcommand{\arraystretch}{1.5}
\renewcommand{\tabcolsep}{0.2cm}

\begin{table}
\caption{\label{tab:HydGround}} \small Comparison of potential energies of the hydrogen ground state computed via damped Bohm SPH simulations with \(N_{S} = 1237\), to the best fit single Gaussian (SG) with width H = 1.33 \(\si{a_{B}}\) and to the exact energy contributions of a 1s wavefunction. Average values of Bohm SPH potentials are calculated from all three runs over 5 snapshots from \(t = 1.36\) to \(1.44 \, \si{fs}\) at \(0.02 \, \si{fs}\) intervals, error given is the standard deviation. All energy values in Hartree units, SG values given to 3 significant figures.
\begin{tabular}[t]{|c|c|c|c|c|}
\hline
Type & $\langle V_{\si{Coul}} \rangle$ & $\langle V_{\si{Bohm}} \rangle$ & $\langle V_{\si{Total}} \rangle$ \\
\hline 
Bohm SPH &  -1.05 $\pm$ 0.01 & 0.59 $\pm$ 0.01 & -0.46 $\pm$ 0.01 \\
SG 1.33 $ \, \si{a_{B}}$ & -0.849 & 0.424 & -0.424 \\
1s &  -1.0 & 0.5 & -0.5 \\
\hline

\end{tabular}
\end{table}

\section{\label{sec:results}Warm Dense Hydrogen Results}

Bohm SPH was used to model a many-body system of spin unpolarised hydrogen at a density of \(n_{e} = 3.006 \, \si{g}/\si{cm^{3}}\) and temperature \(T = 21.54 \, \si{eV}\), corresponding to \(\theta = 1.32\) and \(r_{s} = 1.75 \, \si{a_{B}}\). At these conditions the ion coupling is \(\Gamma_{i} = (Ze)^{2}/(4 \pi \epsilon_{0} a_{i} k_{B} T) = 0.72\) with \(a_{i} =r_{s}\), and the electron plasma period is \(0.203 \, \si{fs}\). The system has 512 protons and 16384 SPH electron particles (\(N_{ppe} = 32\)). Importantly, with the kernel widths dynamically updated according to equation \eqref{eq:denswid} with \(\zeta = 1.3\), the average SPH kernel width \(\overline{h} = 1.16 \, \si{a_{B}}\) is less than the expected screening length of the plasma \(\lambda_{S} = 1.29 \,\si{a_{B}}\) for these parameters. The system is evolved with a time step of \(0.25 \, \si{as}\) in all simulations.

The following simulations have three stages. A first stage of 50 \(\,\si{fs}\) when a thermostat is applied to the ions and the SPH particles remain in NVE to allow them to converge on their centres of mass. A second stage of \(250\, \si{fs}\) when a thermostat is also applied to the electron centres of mass to bring them to the same temperature as the ions. Finally the third stage of 0.7 \(\si{ps}\) where both the ions and the SPH particles are released into a microcanonical ensemble in which trajectory data is collected. We note that for our target density and temperature the exact Fermi-Dirac kinetic energy distribution differs only mildly from a Maxwellian, so we have allowed the electrons to relax into a Maxwellian distribution for the collection of trajectory data. 

\begin{figure}[h]
\includegraphics[width=0.48\textwidth]{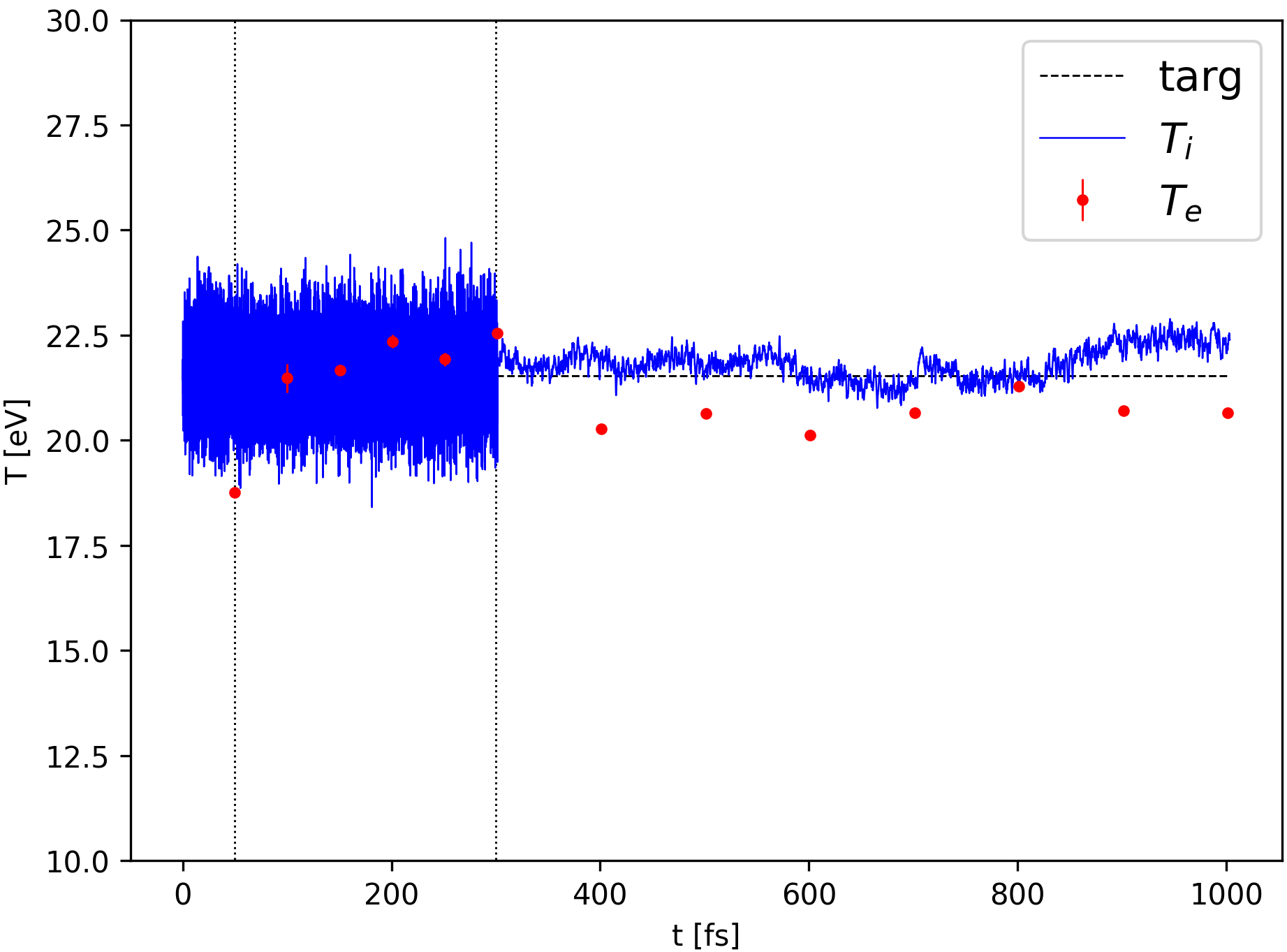}
\caption{\label{fig:Thermalisation} \small Temperature data for run of strongest confinement (\(g = 8.16 \, \si{Ha}/\si{a_{B}}^{2}\)) of Bohm SPH. `targ' corresponds to the target temperature, \(T_{i}\) the ion temperature, and \(T_{e}\) the electron temperature as defined by equation \eqref{eq:e_temp}. Error bars on \(T_{e}\) are the standard deviation of the temperature computed at 25 individual timesteps separated by 1.0 \(\si{as}\) each (centre point is mean). The simulation stage boundaries are indicated by the vertical dotted lines.}
\end{figure}

An example of the thermalisation of the system is shown in figure \ref{fig:Thermalisation}, with the ion temperature \(T_{i}\) and the electron temperature \(T_{e}\) plotted.  For free SPH particles the electronic temperature is simply given by equation \eqref{eq:equipart}. Instead, we are using confining potentials and also a thermostat that operates on the centre of mass velocity of each electron \(\mathbf{V}_{i} = \sum^{N_{ppe}}_{j} \mathbf{v}_{j}/N_{ppe}\). The trajectories of the electron centres of mass are then collected to compute the electronic structure. Hence the centre of mass temperature is the electronic temperature of interest to us

\begin{equation}
\sum_{i}^{N_{e}} \frac{1}{2} m_{e} \langle \mathbf{V}_{i}^{2}\rangle = \frac{3}{2} N_{e} k_{B} T_{e}.
\label{eq:e_temp}
\end{equation}

\begin{figure}[h]
\includegraphics[width=0.48\textwidth]{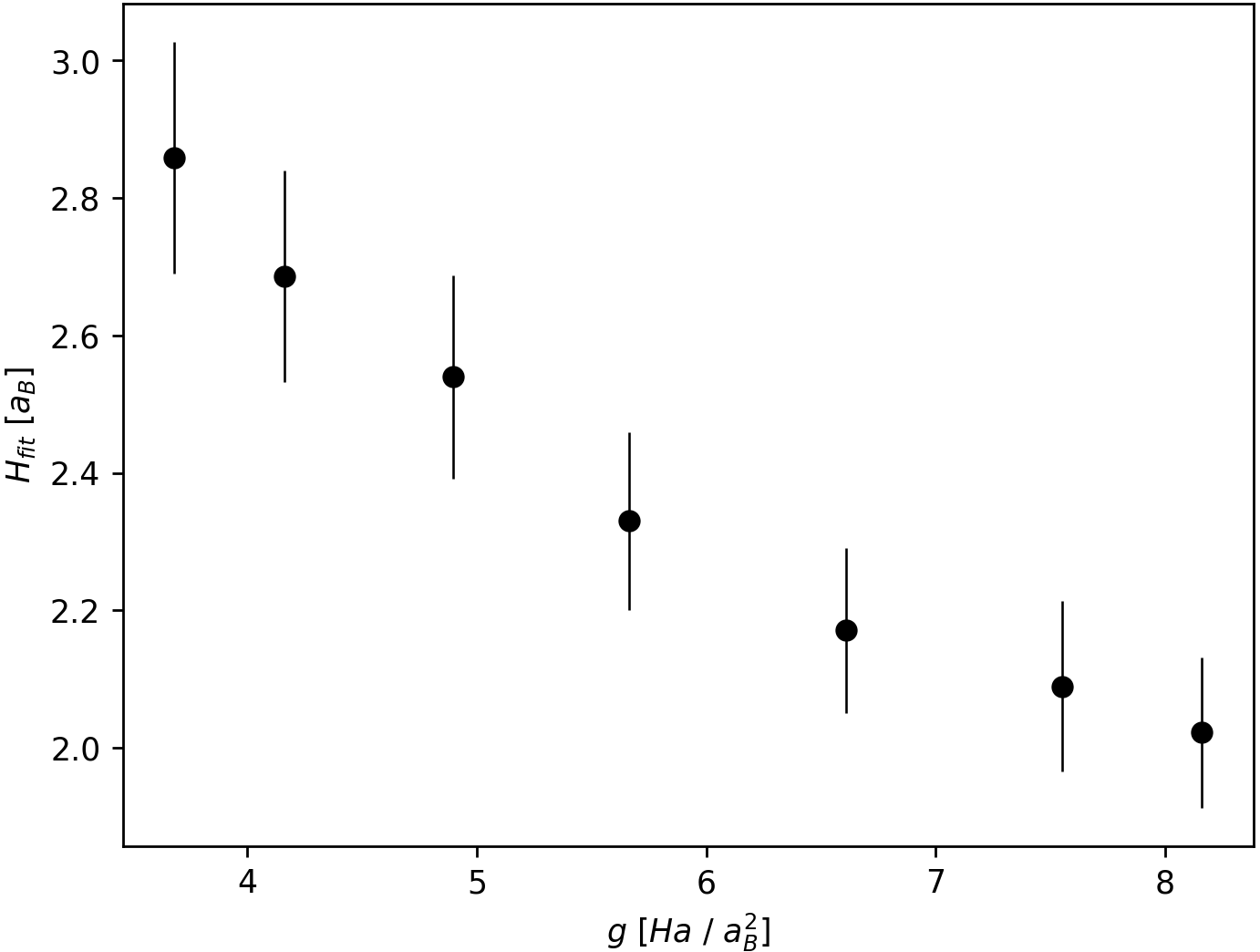}
\caption{\label{fig:KEConfine} \small Mean electron Gaussian widths (\(\pm\) standard deviation) from end of Bohm SPH runs of hydrogen at \(\theta = 1.32\) and \(r_{s} = 1.75 \, \si{a_{B}}\) with different confinement strengths \(g\).}
\end{figure}

\begin{figure}[h]
\includegraphics[width=0.48\textwidth]{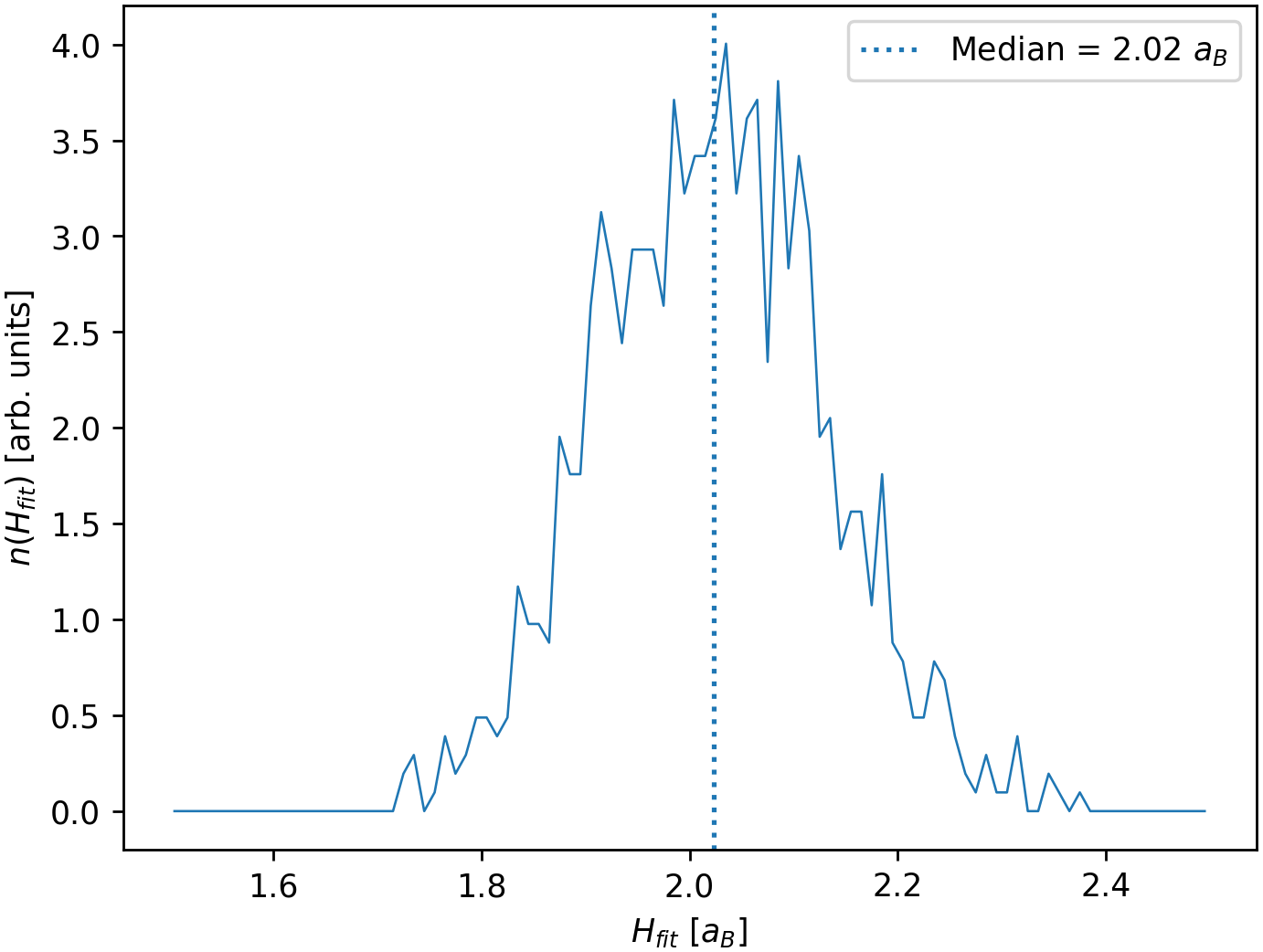}
\caption{\label{fig:EleWidths} \small Fitted electron width distribution from end of Bohm SPH runs with strongest confinement and best agreement to WPMD.}
\end{figure}

We scan values of \(g\) producing electron sizes between roughly 3.0 and 2.0 \(\si{a_{B}}\) (as shown in figure \ref{fig:KEConfine}), calculated by fitting a single Gaussian to the density distribution of SPH particles belonging to the same electron. At each confinement strength we perform two runs with different initial conditions to average the results. The drift in total energy over the 0.7 \(\si{ps}\) of data collection when under the strongest confinement is less than \(5\%\) of the total kinetic energy at release into NVE. An example distribution of the fitted electron sizes in the strongest confinement case is given in figure \ref{fig:EleWidths}. The plateauing trend of mean fitted widths in figure \ref{fig:KEConfine} suggests substantial further contraction of the electron width may not be feasible with our selected SPH parameters. A larger value of \(N_{ppe}\) may allow investigation of smaller electron widths by decreasing the average particle kernel width.

The results are benchmarked against outputs from anisotropic WPMD, in which the root mean square width of the Gaussian wavepackets was \(H_{W} = 1.44 \, \si{a_{B}}\). A key quantity of interest is the dynamic structure factor (DSF), defined for systems in thermodynamic equilibrium as

\begin{eqnarray}
S(\mathbf{k},\omega) = \frac{1}{2\pi N} \int dt \, \exp({i\omega t}) \langle \rho(\mathbf{k},t)\rho(-\mathbf{k},0)\rangle,
\label{eq:DSF}
\end{eqnarray}

\noindent where \(N\) is the number of particles and \(\rho(\mathbf{k},t)\) is the spatial Fourier transform of the time-dependent density \(n(\mathbf{r},t)\). The dynamic structure factor is the power spectrum of the intermediate scattering function \cite{hansen2013theory}

\begin{eqnarray}
F(\mathbf{k},t) = \frac{1}{N} \langle \rho(\mathbf{k},t)\rho(-\mathbf{k},0)\rangle.
\label{eq:ISF}
\end{eqnarray}

The dynamic structure factor describes density fluctuations at wavenumber \(\mathbf{k}\) and frequency \(\omega\), and is an essential link between theory and experiment, with x-ray thomson scattering deployed to diagnose the density and temperature of dense plasmas in the laboratory \cite{fletcher2014observations,poole2022case}, where the experimentally measured x-ray scattering cross section is directly proportional to the total dynamic structure factor of the electrons \cite{glenzer2009x,gregori2009low}. We also examine the static structure factor, calculated via frequency integration of the DSF \(S(\mathbf{k}) = \int d\omega \, S(\mathbf{k},\omega)\), and also related (via Fourier transform) to the pair correlation function. In the following results we assume isotropic and spatially uniform systems such that the structure factors depend only on the magnitude of the wavenumber \(k = |\mathbf{k}|\). 

\begin{figure*}[p]
\includegraphics[width=0.97\textwidth]{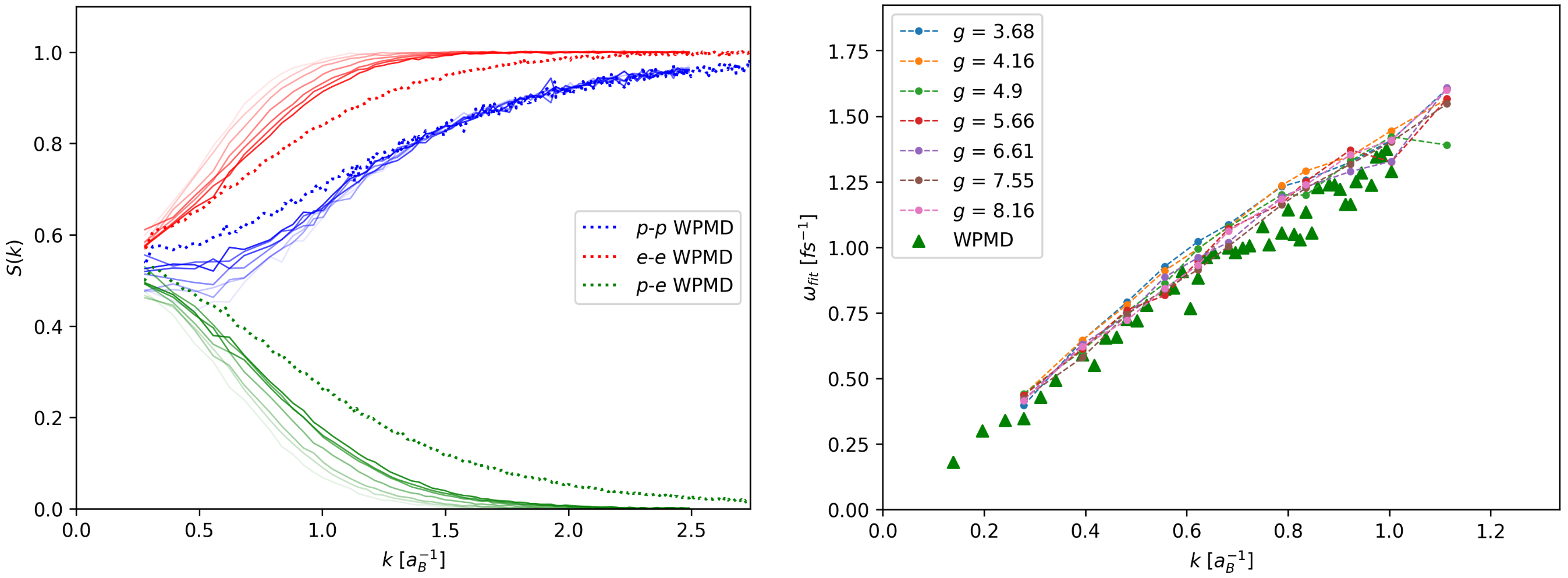}
\caption{\label{fig:gscanStruct}\small Left: Proton - proton and electron - electron static structure factors from Bohm SPH runs with confinement compared to reference calculation from WPMD. Values of \(g\) as in figure \ref{fig:KEConfine} with smallest confinement in lightest shade to strongest confinement in darkest, units \(\si{Ha}/\si{a_{B}}^{2}\). Right: Ion dispersion from Bohm SPH scan of confinement strengths. Frequency plotted is the fitted GCM value for the propagating mode.}
\end{figure*}

\begin{figure*}[p]
\includegraphics[width=1.0\textwidth]{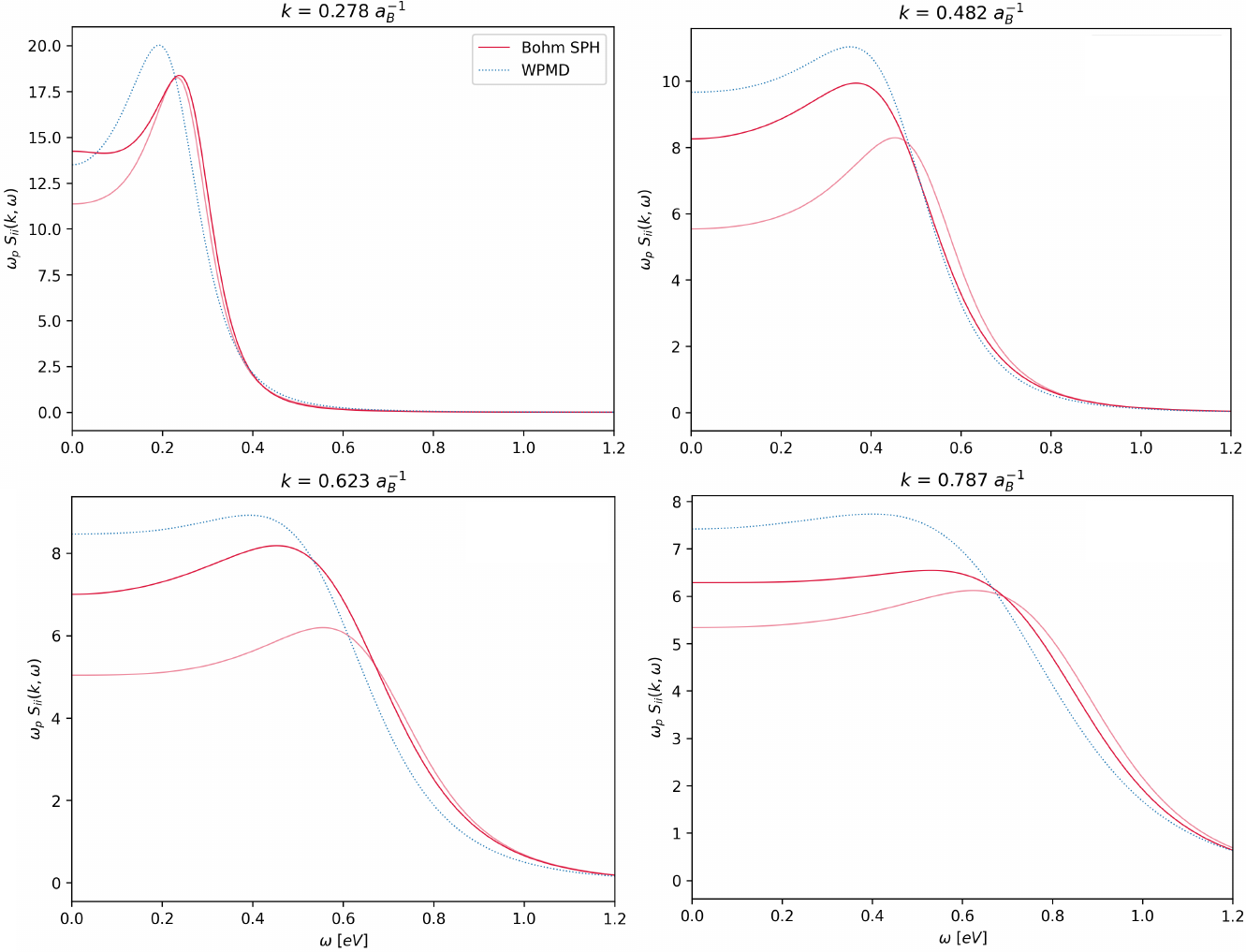}
\caption{\small Ion dynamic structure factors for selected \(k\) modes for strongest (\(g = 8.16 \, \si{Ha}/\si{a_{B}}^{2}\), dark red) and weakest (\(g = 3.68 \, \si{Ha}/\si{a_{B}}^{2}\), light red) confinement. Compared to WPMD outputs (dotted blue).} 
\label{fig:ionDSFs}
\end{figure*}

\begin{figure*}[p]
\includegraphics[width=1.0\textwidth]{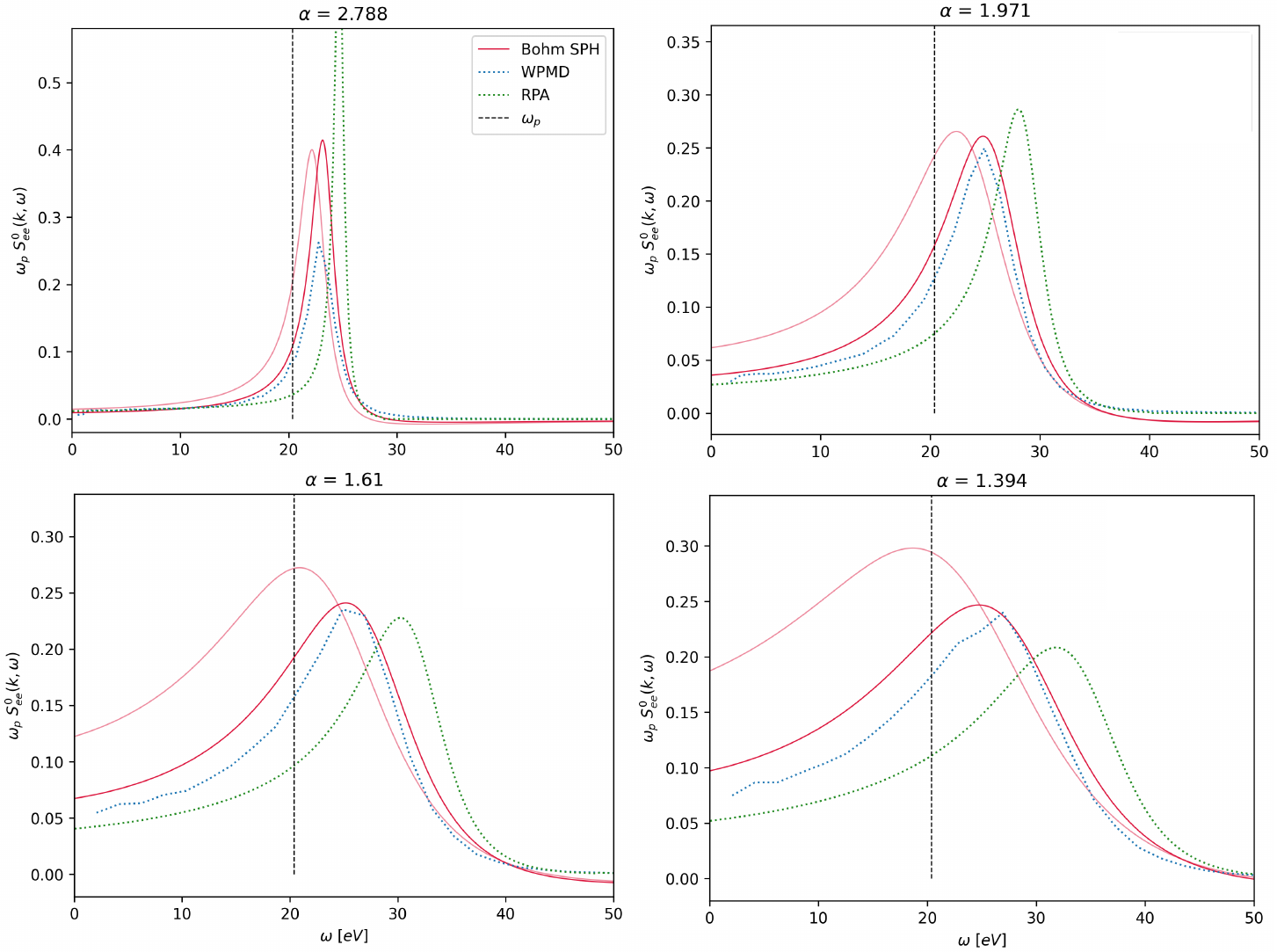}
\caption{\small Free electron dynamic structure factors for collective scattering k values \(\alpha > 1/(k\lambda_{S})\) for strongest (\(g = 8.16 \, \si{Ha}/\si{a_{B}}^{2}\), dark red) and weakest (\(g = 3.68 \, \si{Ha}/\si{a_{B}}^{2}\), light red) confinement. Compared to WPMD (dotted blue) and RPA (dotted green) outputs.}
\label{fig:eleDSFs}
\end{figure*}

\begin{figure*}[p]
\includegraphics[width=1.0\textwidth]{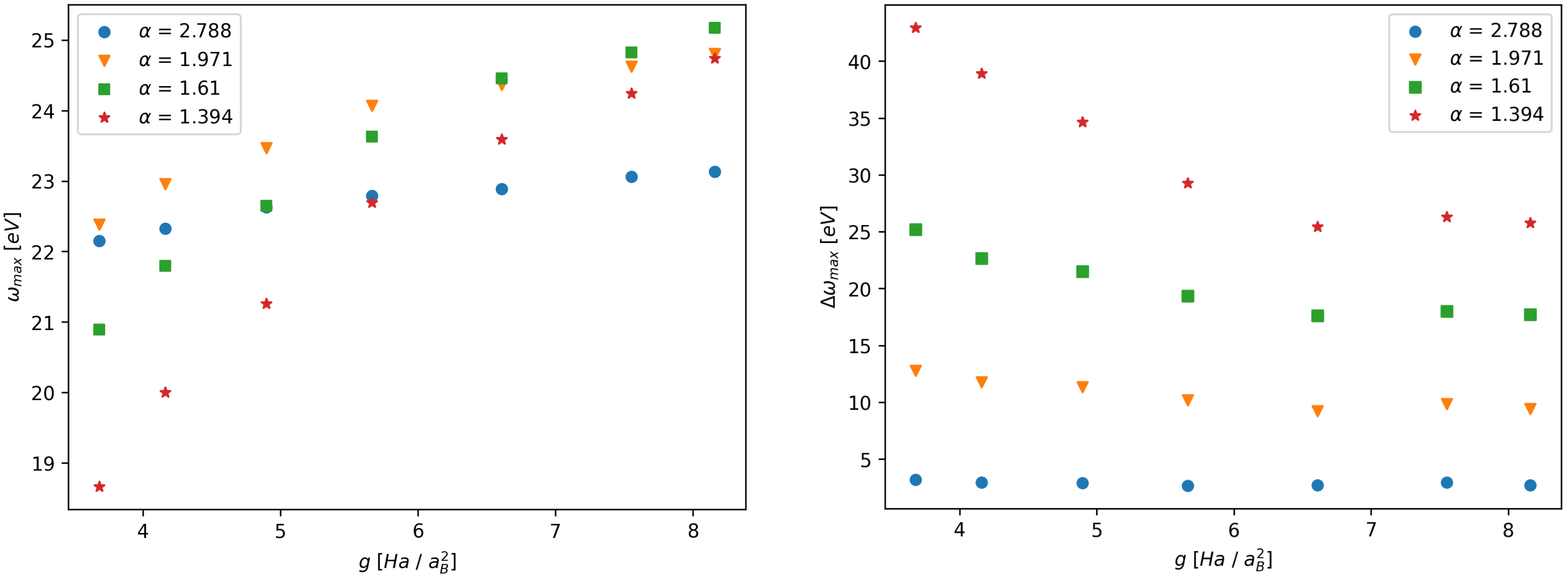}
\caption
{\small } Maximum value (left) and  FWHM of plasmon (right). Computed for collective scattering k values \(\alpha > 1/(k\lambda_{S})\) across all confinement strengths \(g\) sampled.
\label{fig:plasmonconverge}
\end{figure*}

When presenting dynamic structure data from Bohm SPH, we have employed the generalized collective modes (GCM) approach, as described in Ref. \cite{wax2013effective} and deployed in analysis of ionic modes in Ref. \cite{schorner2022extending}. We perform the fits of the intermediate scattering functions using one propagating and one diffusive mode, then used to calculate associated dynamic structure factors \(S(k,\omega)\).

Figure \ref{fig:gscanStruct}a demonstrates that the Bohm SPH static structure calculations have improved agreement with the WPMD calculation as the strength of confinement is increased. Unsurprisingly, the ion-electron and electron-electron structure factors are more sensitive to the strength of confinement.  Even in the case of the weakest confinement however, the ion structure agrees reasonably well with WPMD, and the extrapolated electron and ion structure values at \(S(k=0)\), related to the compressibility \cite{gregori2007derivation,pines2018theory}, are similar to the WPMD estimates. We ascribe the difference in static structure observed between Bohm SPH and WPMD to be primarily due to different electron sizes, which strongly affect the screening of the plasma. The strongest confinement case of Bohm SPH achieves an average electron width of \(H_{fit} = 2.02 \, \si{a_{B}}\), still larger than the root mean squared width of the WPMD output of \(H_{W} = 1.44 \, \si{a_{B}}\). 

The ion dispersion is relatively insensitive to the confinement strength, as shown in figure \ref{fig:gscanStruct}b. If we also examine the ion dynamic structure factor, as in figure \ref{fig:ionDSFs}, we can see relatively good agreement between Bohm SPH and WPMD, with some differences in the strength of the diffusive mode.

Using the centre of mass coordinates of each electron recorded over the simulation, and treating them as point particles, we also compute the electron dynamic structure. A commonly used decomposition of the electron dynamic structure factor is given by Chihara \cite{chihara1987difference,chihara2000interaction} 

\begin{equation}
\begin{split}
S_{ee}(\mathbf{k},\omega) = |f(\mathbf{k}) + n(\mathbf{k})|^{2}S_{ii}(\mathbf{k},\omega) + S_{ee}^{0}(\mathbf{k},\omega) \\ + S_{bf}(\mathbf{k},\omega)
\label{eq:chihara}
\end{split}
\end{equation}

\noindent where \(f(\mathbf{k})\) is the unscreened bound electron form factor, \(n(\mathbf{k})\) the screening cloud form factor, \(S_{ii}(\mathbf{k},\omega)\) the ion - ion structure factor,  \(S_{ee}^{0}(\mathbf{k},\omega)\) the free electron structure factor, and \(S_{bf}(\mathbf{k},\omega)\) a scattering contribution from bound-free transitions. In our simulation of ionized hydrogen, with no contribution from \(f(\mathbf{k})\) or \(S_{bf}(\mathbf{k},\omega)\) we have access to both \(S_{ii}(\mathbf{k},\omega)\) and \(S_{ee}(\mathbf{k},\omega)\). Comparison of the intermediate scattering functions \(F_{ee}(\mathbf{k},t)\) and \(F_{ii}(\mathbf{k},t)\) enables calculation of the screening cloud form factor \(n(\mathbf{k})\) and by extension, isolation of the free electron structure factor \(S_{ee}^{0}(\mathbf{k},\omega)\) \cite{svensson2024modeling}. The screening cloud \(n(\mathbf{k})\) can also be computed by comparing the proton-proton and proton-electron static structure factors \cite{chihara1987difference} via \(S_{pe}(\mathbf{k}) = n(\mathbf{k})S_{pp}(\mathbf{k})\) in the case of hydrogen. Here, we compute \(n(k)\) (isotropic) by minimising the loss

\begin{eqnarray}
L = \int dt \, \big[F_{ee}(k,t) - (n(k))^{2}F_{ii}(k,t)\big]^{2}.
\label{eq:nkloss}
\end{eqnarray}

For small values of \(k\) in the collective regime \( \alpha =  1/k\lambda_{S} > 1\), we apply the GCM fitting procedure as before with one propagating and one relaxing mode. In addition, we apply a detailed balance correction, as in Ref. \cite{gregori2009low}, of the form \(\beta \hbar \omega/(1 - e^{-\beta \hbar \omega})\).

The outputs are plotted in figure \ref{fig:eleDSFs}, and they compare favourably with outputs from WPMD, computed via direct fourier transform of the truncated intermediate scattering function and which apply the same detailed balance correction. In the electron dynamic structure factors the effect of confinement is more prominent. Both the position of the plasmon peak and the value of \(S_{ee}^{0}(k,\omega = 0)\) agree more closely with WPMD in the strongly confined case than weakly. The weakly confined case consistently underpredicts the plasmon frequency and overestimates \(S_{ee}^{0}(k,\omega = 0)\), associated with the electron diffusivity, when comparing to WPMD. Figure \ref{fig:plasmonconverge} shows how the plasmon frequency and its width trend with increasing confinement strength. In the collective \(\alpha >1 \) regime, the values reasonably converge by the strongest confinement case. With a more pronounced dependence on confinement strength at shorter length scales (smaller \(\alpha\)), we see how the achieved electron size determines the resolvable electron dynamics.

The electron dynamic structure outputs are also compared to the predictions of the Random Phase Approximation \cite{pines1952collective,pines2018theory}, which applies when the interparticle interactions are weak. We note that the numerical outputs for the plasmon (strong confinement Bohm SPH and WPMD) at the investigated k modes predict a lower peak frequency and a slightly broader plasmon. A similar effect has been reported in previous work investigating the impact of exchange-correlation as well as ion collisions on plasmon dispersion \cite{dornheim2018ab,fortmann2010influence}.

\section{\label{sec:conclusion}Conclusions and Future Work}

We have presented a new scheme for the simulation of WDM. Advantages of the methodology are its non-adiabatic treatment of ion - electron interactions with explicit electron dynamics, a many-body calculation of the Bohm potential, the ability to model arbitrary electron shapes, tunable resolution, and computational scalability. After conservations tests, the Bohm and Coulomb implementations of the code were validated by single particle tests of the quantum harmonic oscillator ground state and the hydrogen 1s wavefunction.

The non-adiabatic treatment of the ion-electron interaction when using more SPH particles than electrons present in the system motivates use of a confining potential to localise individual electrons, whose centre of mass velocity can be operated upon by a thermostat to achieve an appropriate distribution. 

Bohm SPH was used to simulate a warm dense hydrogen system at \(\theta = 1.32\) and \(r_{s} = 1.75 \, \si{a_{B}}\) and compared to outputs from anisotropic WPMD, scanning a range of confinement strengths. In particular, the electron dynamic structure factors of the strongest confinement case agreed well with outputs from WPMD in the collective regime. Comparison of static structure outputs were also encouraging while indicating that a smaller electron size in Bohm SPH would improve agreement with WPMD. More broadly, comparisons of Bohm SPH outputs for the static and dynamic structure factors when scanning the confinement strength show how the electron size affects screening within the plasma.

Several aspects of the methodology might be improved or generalised in future versions. The treatment here is based upon Gaussian kernels, but generalising the Coulomb interaction to any kernel function would allow use of kernels with better compact support, reducing computational cost. Additionally, the Coulomb interaction could be extended to include the interaction between SPH particles and ions with core electrons via a pseudopotential. Although the fixed-point iterator used in this work to self consistently calculate SPH densities and scale lengths performed well, the Newton-Raphson root finding procedure would be preferable. Finally, further forms of density derivative used in computing the Bohm pressure tensor, as compared in section \ref{sec:conserve}, could be explored.

\section{\label{sec:acknowledgements}Acknowledgements}

Computing resources were provided by STFC Scientific Computing Department’s SCARF cluster. We are grateful to the anonymous referee whose comments improved this paper. T.C. would like to acknowledge useful discussions with Thomas Gawne, Sam Azadi, and Thomas White prior to and during the production of Bohm SPH. T.C. and S.M.V. acknowledge support from the Royal Society and from the UK EPSRC grant EP/W010097/1. P.S. acknowledges funding from the Oxford Physics Endowment for Graduates (OXPEG). P.S., D.P., S.M.V. and G.G. acknowledge support from AWE-NST via the Oxford Centre for High Energy Density Science (OxCHEDS).

\section{\label{sec:appendix}Appendix: Momentum Conservation}

\begin{figure}[h]
\includegraphics[width=0.48\textwidth]{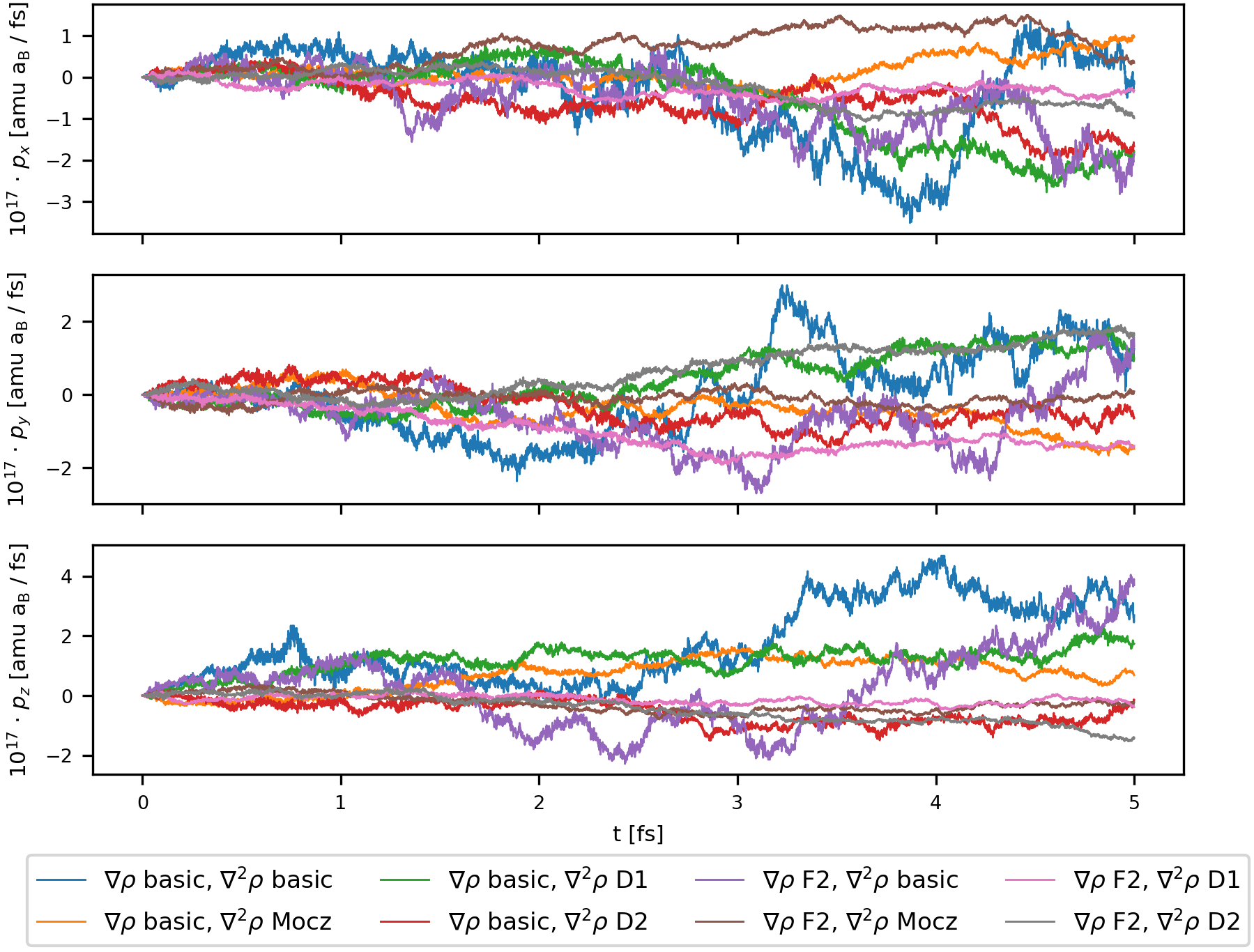}
\caption{\label{fig:bohm_mom}\small Total momentum components of a Bohm-only system with different derivative combinations. Y axes multiplied by a factor of \(10^{17}\).}
\end{figure}

\begin{figure}[h]
\includegraphics[width=0.48\textwidth]{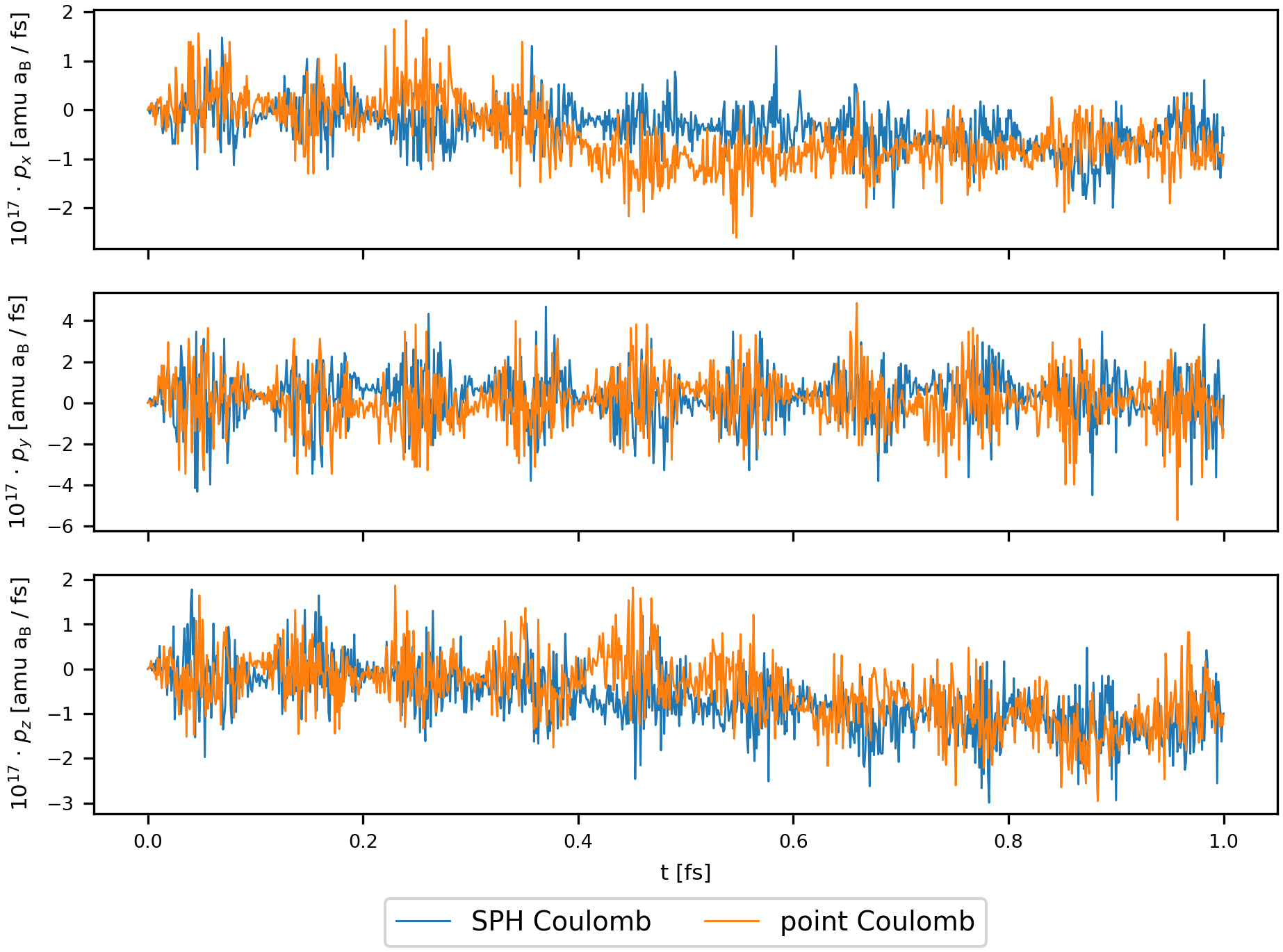}
\caption{\label{fig:coul_mom}\small Total momentum components of a test OCP Coulomb system comparing standard point and SPH Coulomb interactions. Y axes multiplied by a factor of \(10^{17}\).}
\end{figure}

\providecommand{\noopsort}[1]{}\providecommand{\singleletter}[1]{#1}%

\end{document}